\documentclass[times,sort&compress,3p]{elsarticle}
\usepackage{amsmath, amssymb, enumerate, amsthm}
\usepackage[colorlinks,citecolor=blue,urlcolor=blue]{hyperref}
\usepackage{booktabs}
\usepackage[plain,noend]{algorithm2e}
\usepackage{float}
\usepackage{tabularx}
\usepackage{times,bm}
\usepackage{multirow}
\newcommand{\CG}[1]{\textcolor{black}{#1}}
\newtheorem{theorem}{Theorem}
\newtheorem{lemma}{Lemma}

\newtheorem{corollary}{Corollary}
\DeclareMathOperator*{\dmin}{\underline{m}}
\newcommand{\sstar}{\ensuremath{s^{\star}}}
\renewcommand{\widehat}{\hat}
\usepackage{lipsum}
\makeatletter
\def\ps@pprintTitle{%
	\let\@oddhead\@empty
	\let\@evenhead\@empty
	\def\@oddfoot{}%
	\let\@evenfoot\@oddfoot}
\makeatother
\begin{document}

\begin{frontmatter}

\title{Feature screening in large scale cluster analysis}

\author[mymainaddress]{Trambak Banerjee}
\author[mymainaddress]{Gourab Mukherjee}
\author[mysecondaryaddress]{Peter Radchenko\corref{mycorrespondingauthor}}
\cortext[mycorrespondingauthor]{Corresponding author}
\ead{peter.radchenko@sydney.edu.au}

\address[mymainaddress]{University of Southern California, Los Angeles, CA 90007, USA}
\address[mysecondaryaddress]{University of Sydney, NSW 2006, Australia}

\begin{abstract}
 We propose a novel methodology for feature screening in the clustering of massive datasets, in which both the number of features and the number of observations can potentially be very large.  Taking advantage of a fusion penalization based convex clustering criterion, we propose a highly scalable screening procedure that efficiently discards non-informative features by first computing a clustering score corresponding to the clustering tree constructed for each feature, and then thresholding the resulting values.  We provide theoretical support for our approach by establishing uniform non-asymptotic bounds on the clustering scores of the ``noise'' features.  These bounds imply perfect screening of non-informative features with high probability and are derived via careful analysis of the empirical processes corresponding to the clustering trees that are constructed for each of the features by the associated clustering procedure.  Through extensive simulation experiments, we compare the performance of our proposed method with other screening approaches popularly used in cluster analysis and obtain encouraging results.  We demonstrate empirically that our method is applicable to cluster analysis of big datasets arising in single-cell gene expression studies.
\end{abstract}


\end{frontmatter}

\section{Introduction}
\label{sec1}
We consider the problem of feature screening in large scale cluster analysis.
Clustering is one of the most popular unsupervised classification techniques; \CG{it} is widely used in a myriad of statistical applications for stratification and sub-population identification \cite{farcomeni2016robust,hartigan1979algorithm,rousseeuw1990finding}. In recent years, due to massive advancements in the modern data collection and assimilation techniques, very big datasets, with both a large number of observations and a large number of features, have been generated with increasing frequency.  Classical clustering methods (see, \CG{e.g., Chap.} 14 of \citet{friedman2001elements}) are either computationally challenging or ineffective for conducting segmentation analysis of such massive modern data.  In many scientific applications, when the dimension of the data is very high,  most of the coordinates (i.e., features) contain very little information regarding the grouping structure.  Classical clustering methods, which do not reduce the dimension of the data, suffer, because the agglomerative effects of the large number of ``noise'' features conceal important clustering information available in a relatively smaller number of ``signal'' features.

Recently developed clustering algorithms, which exploit the underlying sparseness, are effective in dealing with high-dimensional data; see, e.g., \cite{arias2014detection,chan2012using, jin2016influential,witten2012framework}. \CG{We propose herein} a scalable computationally efficient approach, entitled COSCI (COnvex Screening for Cluster Information), that can efficiently weed out the features that are non-informative for clustering.  As a \CG{nonparametric} approach, COSCI  has competitive advantages over the popular Gaussian mixture based parametric techniques \citep{arias2014detection,azizyan2013minimax}. Unlike the \CG{nonparametric} density estimation based screening techniques, COSCI is very scalable, and can successfully handle datasets with more than one million observations. Our proposed procedure discards non-informative features by first computing a clustering score for the clustering tree constructed for each feature, and then thresholding the resulting values.  We provide the theoretical motivation for our approach by establishing uniform non-asymptotic bounds on the clustering scores of the noise features.

Our theoretical results are a significant extension of the univariate results in \citet{radchenko2014consistent}. \CG{We rely herein} on a more careful analysis of the empirical processes corresponding to the clustering trees constructed for each feature by the associated clustering procedure.  We derive a stronger tail probability bound for the clustering score of each feature, which then allows us to establish a uniform bound for all the clustering scores of the  non-informative features in the high-dimensional setting, where the number of such features can be extremely large.  Using this uniform bound, we infer that under mild regularity conditions on the population distribution, the proposed COSCI algorithm will discard the non-informative features and perfectly select the informative features with very high probability.

\CG{High-dimensional} datasets arising in modern biology, econometrics, engineering, text mining and signal processing  \citep{james2013introduction} require a significant degree of feature screening for subsequent application of a clustering algorithm.  We illustrate the applicability of COSCI for making scientific discoveries through the analysis of single cell biological data. Emerging technologies, such as single-cell mass cytometry \citep{Bendall11}, next-generation sequencing \citep{liu2012comparison} and micro-fluidic methods \citep{dalerba2011single,white2011high}, have recently enabled us to collect gene and protein expression information for each cell \citep{wang2010single}. The resulting datasets, which are not only very high-dimensional but also contain a large number of cellular observations, serve as invaluable resources for the characterization of the cellular hierarchy in multi-cellular organisms.  Often, the biological question associated with these datasets is the identification of homogeneous cellular sub-populations based on the differential expression patterns of the genes.  The composition of these sub-populations is subsequently analyzed to detect interesting structures.

Recently, several algorithms such as viSNE \citep{amir2013visne}, Wanderlust \citep{bendall2014single}, ECLAIR \citep{giecold2016robust}, SPADE \citep{qiu2011extracting}, SLIDE \citep{sen2015single} and Scaffold maps \citep{spitzer2015interactive} have been developed for conducting such sub-population analysis for single cell data.
Most of these methods conduct either dimension reduction through PCA related methods or handle large sample sizes via down-sampling.  While these algorithms are widely used, they lack appropriate mathematical guarantees to show that the resulting sub-populations are not due to random fluctuations and will be reproducible across datasets generated from experiments conducted under similar conditions. Furthermore, the use of techniques such as PCA to reduce dimensionality in these settings can be called into question \citep{chang1983using} because  (a) the inferred sub-populations may not be sparse in the expression patterns of the genes, in which case PCA has been theoretically proven to produce inconsistent results \citep{johnstone2012consistency}; (b) there is no guarantee that cluster information is aligned in the direction of maximum variance.

As a motivational example, consider the problem of cellular sub-population detection in a single-cell RNAseq data analyzed in \citet{giecold2016robust} (henceforth referred to as G16). This massive dataset holds the expression levels of $p=8716$ genes for $n=2730$ mice bone marrow cells \citep{paul2015transcriptional}. \CG{One} of the key scientific objectives is to infer the lineage pattern of the identified sub-populations based on~$33$ lineage markers. If the sub-populations differ with respect to a relatively small subset of all the genes considered, then, from the statistical perspective, the problem reduces to screening out the non-informative features from the data consisting of vectors \CG{$X_1, \ldots, X_p \in \mathbb{R}^{n}$} and identifying a subset of features that retains the cluster information. As such, identifying the best subset of the genes with respect to the cluster information is an important statistical endeavor with critical biological implications; \CG{see \cite{witten2012framework} and the references therein}.

\begin{figure}[t!]
	\centering
	\includegraphics[height=25pc,width=.9\textwidth]{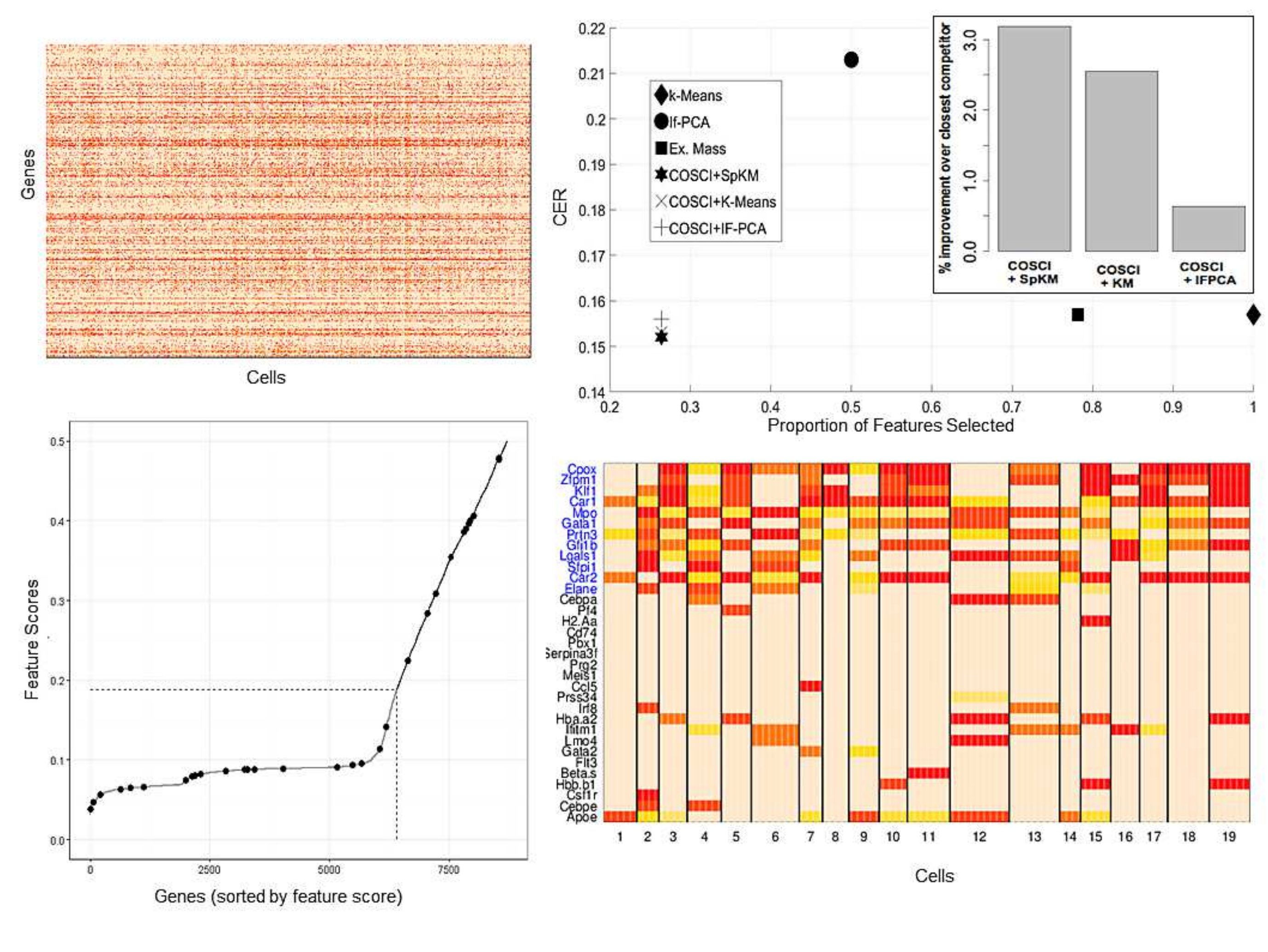}
	\caption{Application of COSCI to RNA-Seq data of G16: At top left we have the heatmap of the $8716\times 2730$ expression matrix, where red denotes high expression and bisque stands for low expression. A priori we know that there are~$19$ sub-populations among the cells. The goal is to detect these sub-populations and study their composition with respect to the~$33$ lineage markers. The plot at bottom left shows the distribution of feature scores~$S_j$.  The dashed horizontal line is the screening threshold of~$0.188$, which is chosen by COSCI.  All the features with scores above the threshold are selected. The black dots are the 33 lineage markers. The heatmap at the bottom right shows the composition of the~$19$ cellular sub-populations among the~$33$ lineage markers. The~$12$ lineage markers in blue are the selected ones. At top right is the plot of the error rate ($y$-axis) vs. the proportion of features selected ($x$-axis) for different methods. COSCI selects the fewest features and applying $k$-means or Sparse $k$-means to the COSCI selected features returns a smaller error rate.}	\label{fig1}
\end{figure}

Unlike the methods that use downsampling, COSCI can accommodate sample sizes on the order of $10^6$.  COSCI first produces a score, $S_j\in(0,0.5]$, for each feature, which reflects its relative importance for clustering, and then screens out the features with lower scores.   When applied to the aforementioned RNASeq data, COSCI orders the scores of the~$8716$ genes and selects the top $2304$ genes, which include~$12$ of the~$33$ lineage markers (see Figure~\ref{fig1}, bottom left; the selected genes are highlighted in solid black and the black dots are the~$33$ lineage markers). The heatmap (Figure~\ref{fig1}, bottom right) of the expressions for the~$19$ sub-populations detected via $k$-means on the COSCI selected features shows that the inferred sub-populations differ significantly across the~$12$ selected lineage markers. After applying $k$-means to detect the sub-populations on the~$2304$ genes selected by COSCI we got a misclassification error (computed as CER = $1-$ Rand Index, \cite{chipman2006hybrid,rand1971objective}) of approximately~$0.15$. This error is significantly smaller than several other methods that are widely used for such clustering problems especially when the proportion of selected features is taken into account (see Figure~\ref{fig1}, top right).  We revisit this example with more details in Section~\ref{rnasec}.

\subsection{Connections to related statistical literature}

Within the statistical literature, a number of recently proposed clustering approaches execute feature screening as the first step and then rely on conventional clustering techniques, such as $k$-means, to cluster the remaining data. For example, \citet{chan2012using} proposed a \CG{nonparametric} feature screening method that is based on coordinate-wise \CG{excess mass} tests \cite{cheng1998calibrating}. They rank the features using the values of the corresponding test statistic. Feature selection then follows by identifying a kink in the plot of the \CG{within-cluster} sum of squares versus the number of identified clusters. \citet{witten2012framework} proposed the sparse $k$-means and sparse hierarchical clustering approaches, which employ $k$-means and hierarchical clustering, respectively, on a feature weighted dissimilarity matrix, where the weights are encouraged to be sparse.  Their method is largely inspired by the popular COSA algorithm  of \cite{friedman2004clustering} and is more adept at sparse clustering.  Recently, \citet{arias2016simple} proposed Sparse Alternate Sum (SAS) clustering, which uses a hill-climbing approach to solve the sparse $k$-means optimization problem.

On the parametric side, several \CG{model-based} clustering approaches have been introduced \cite{pan2007penalized,wang2008variable,xie2008penalized}. These techniques typically maximize a penalized likelihood under a Gaussian mixture model, where the penalization serves the purpose of implicit feature selection. \citet{jin2015phase} and \citet{jin2016influential} propose IF-PCA, which is a \CG{two-step} clustering method --- the first step conducts coordinate wise feature selection, and the second step performs $k$-means clustering on the matrix of left singular vectors of the selected features. The feature selection step uses the \CG{Kolmogorov--Smirnov} test for normality to rank the features, followed by the use of the Higher Criticism (HC) \cite{donoho2004higher,donoho2008higher} functional to finally select the features.  Theoretical properties of clustering algorithms that combine feature selection with clustering have also been recently studied. For example, \citet{azizyan2013minimax} provide information theoretic bounds on clustering accuracy of the \CG{high-dimensional} Gaussian mixtures, while \citet{arias2014detection} establish minimax rates for the problems of mixture detection and feature selection under the sparsity assumption.

Our work is closer to the approaches of \cite{chan2012using}, \cite{jin2016influential} and \cite{witten2012framework}, where the objective is to screen out the noise features.  We analyze the problem of feature screening in large scale clustering and propose COSCI --- a novel  computationally efficient screening procedure with strong theoretical motivation.  COSCI uses a \CG{nonparametric} approach to rank order the features by their clustering leverage. In this respect, it differs from the recently proposed screening techniques, such as IF-PCA, which rely on a parametric family as a point of reference to gauge feature strength for clustering.

\subsection{Organization of the paper}

In Section~\ref{sec2-intro} we present and discuss our screening methodology.  More specifically,  Algorithm~1 provides the details of the implementation, while Section~\ref{sec2-theory} contains the main theoretical motivation and associated results. Section~\ref{sec3-est} provides two approaches that aid the selection of the screening threshold. In Section~\ref{sec4} we conduct a detailed empirical analysis of our approach using both simulated data and real data from microarray experiments. Section~\ref{sec5} concludes the paper. Proofs and additional technical details are relegated to the Appendix.

\section{Methodology and main results}
\label{sec2-intro}

We consider the problem of clustering~$n$ observations based on~$p$ features in the setting where most of the features contain no clustering information.  Noting that these ``noise'' features have  unimodal marginal distributions (which may differ across the features), we develop a univariate approach, which, based on the sample observations, evaluates whether the true underlying density is unimodal. The theoretical and empirical results provided in Sections~\ref{sec2-theory} and~\ref{sec4} demonstrate that our approach is successful at screening out the noise and identifying the signal features even in very high-dimensional scenarios.

The key ingredients of our proposed COSCI methodology are a univariate merging algorithm, which constructs a clustering tree for each of the features, and a screening of the merge sizes, which identifies the signal features as those with at least one sufficiently large merge.  We discuss each of these components in detail below.

\medskip
\noindent\textit{Univariate merging algorithm.} With the goal of checking unimodality for each of the feature coordinates, we consider the following univariate optimization problem:
\begin{equation}
\label{eq:2}
\min_{c_1,\ldots,c_n \in \mathbb{R}}\sum_{i=1}^{n}(x_{i}-c_{i})^{2}+\lambda\sum_{1\le k<\ell \le n}|c_{k}-c_{\ell}|.
\end{equation}
It is based on the observations~$x_i$ and corresponds to minimizing the \CG{within-cluster} sum of squares under  constraints on the \CG{$L_1$-distance} between the cluster centroids~$c_k$. Here~$\lambda$ is a non-negative penalty weight.  The convexity of the objective function in~\eqref{eq:2} has been exploited to develop algorithms for efficiently producing the path of solutions as a function of the penalty weight \cite{hocking2011clusterpath,hoefling2010path,radchenko2014consistent}. Clustering algorithms based on fusion penalization of this type have become very popular in large scale clustering \cite{chi2015splitting,hocking2011clusterpath,radchenko2014consistent,tan2015statistical,zhu2014convex}  and regression analysis \cite{Bondell-08,ke-13,shen-10,shen-12}.

The entire path of solutions corresponding to the objective criterion \eqref{eq:2} can be found by a simple merge algorithm in $O(n\ln n)$ operations. Starting with~$n$ observations in~$n$ clusters, we sequentially merge the nearest (in terms of the weighted distance as shown in Algorithm~1) adjacent centroids until we are left with just one cluster in the end.

\medskip
\noindent\textit{Merge sizes.} Given a merge of sub-clusters $\mathcal{C}_{1}$ and $\mathcal{C}_{2}$, we calculate its size, which we define as
\begin{equation}
\label{eq:3}
\alpha=n^{-1}\min (|\mathcal{C}_{1}|,|\mathcal{C}_{2}|) \,{\mathbf 1} \{n^{-1} (|\mathcal{C}_{1}|+|\mathcal{C}_{2}| )\ge0.5 \},
\end{equation}
where $|\cdot|$ denotes the cardinality of a set.  The above thresholding of the mass after merge ensures that we only identify a merge as big if it results in a significantly large cluster.  This protects us from the risk of discovering potential big merges on smaller fragments of the sample, where the nature of these merges can be very fragile due to sampling fluctuations.

The fundamental working principle of our proposed COSCI approach rests on the following property of the merge sizes: if $x_1,\ldots,x_n$ are indeed generated by a non-informative density, then the sample merges $\alpha_1, \ldots, \alpha_{n-1}$  will  be uniformly small for sufficiently large~$n$. A multivariate version of this property is formalized in Theorem~\ref{thm1} below. \CG{In contrast}, if the underlying distribution contains a moderate amount of cluster information, then  the merge sequence will have at least one merge that is \textit{big}.  The last fact is illustrated by Theorem~1 in \cite{radchenko2014consistent}; see also the discussion above Corollary~\ref{cor1}. Based on these properties of the merge sizes, we conduct the following screening procedure.

\medskip
\noindent\textit{Screening the merges.} Given a pre-defined threshold $\alpha_0$,  we flag the feature as potential ``signal'', if there exists a corresponding merge, say the \CG{$i$th} merge, such that $\alpha_i\ge\alpha_{0}$.

Our proposed methodology is formalized in Algorithm~1 below.

\bigskip
\noindent
\textbf{Algorithm 1:} COSCI procedure for feature screening

\medskip
\noindent
INPUT: Data matrix ${X}_{n\times p}$ and tuning parameter $\alpha_0$. \newline
FOR each $j\in\{1,\ldots,p\}$:\newline
INITIALIZE:\newline
\indent Sort data in ascending order and store them as $x = \{x_1,\ldots,x_n\}$.\newline
\indent Set $k$, the number of clusters, equal to n. For each $r \in \{1,\ldots, n\}$, set $c_r = \{x_i\}$.\newline
REPEAT:\newline
\indent \quad Find consecutive adjacent centroid distances: $d(r,r + 1)\leftarrow (c_{r+1}-c_{r})/(|c_r| + |c_{r+1}|)$.\newline
\indent \quad Find clusters with minimum merging distance: $r^* \leftarrow \arg \min_r d(r, r + 1)$.\newline
\indent \quad Merge clusters $r^*$, $r^*+1$, re-label remaining clusters and set $k\leftarrow k-1$.\\[2pt]
\indent\quad Find the merge size, $\alpha_{n-k}^j$, using equation  \eqref{eq:3}.\newline
UNTIL $k = 1$\newline
STORE: Clustering score: $S_j = \max_{1\le k\le n-1}\alpha_k^j$\newline
FEATURE SCREENING: $\widehat{\mathcal{I}}_{S}= \{j:S_j\ge\alpha_0 \}$\\[5pt]

See  \ref{sec:appe} for a detailed description of the computational steps involved in \CG{Algorithm~1}.  We note that a significant gain in computational time is achievable via a parallel implementation of the top \textit{for} loop in Algorithm~1 that runs across the $p$ features.

\subsection{Theoretical support: Perfect screening property}
\label{sec2-theory}

Let $\mathcal{I}_S$ and $\mathcal{I}_N$ be the index sets corresponding to the ``signal'' and the ``noise'' features, respectively.  Define $p_S=|\mathcal{I}_{S}|$ and $p_N=|\mathcal{I}_{N}|$.  Given feature~$j$, we write $S_{j}(\tau)$ for the corresponding largest merge size, computed the same way as $S_j$ in Algorithm~1, but under the restriction that the midpoint between the two merged sub-clusters lies between the \CG{sample quantiles of order $100 \times \tau$ and $ 100 \times (1- \tau )$, where $\tau$ is an arbitrarily small but positive number.}

Theorem \ref{thm1}, stated below, establishes a uniform non-asymptotic bound on the merge sizes that are produced when our procedure is applied to the noise features. The regularity conditions, C1 and C2, are imposed on the family of marginal distributions of the noise features, and are fairly mild.  In particular, they are satisfied for location-scale families of unimodal differentiable densities with a finite first moment.  The proof of Theorem~\ref{thm1} is provided in~\ref{sec:appa}. \CG{In what follows, $a \vee b = \max(a,b)$ and $a \wedge b = \min (a,b)$, for any $a, b \in \mathbb{R}$.}

	\begin{theorem}
	Suppose that regularity conditions C1 and C2, stated in~\ref{sec:appa}, are satisfied.   For each $\tau>0$ there exist positive constants $c_1$, $c_2$, $b$ and~$\kappa$, whose choice does not depend on either~$n$ or~$p_N$, such that, as long as $p_N\le \exp(\kappa n)$, inequalities
	\begin{equation*}
	\max_{j\in\mathcal{I}_N} S_{j}(\tau) \le b\,\dfrac{\ln (p_N\vee n)}{n}
	\end{equation*}
	hold with (high) probability bounded below by $1-c_1p_N^{-c_2}$.
	\label{thm1}
\end{theorem}
The above theorem provides the theoretical justification for the screening step in our proposed procedure. The result is non-asymptotic, and the proof involves careful analysis of the empirical process associated with the merging algorithm for each feature. Under a mild restriction on the number of features relative to the sample size, the above theorem ensures that the clustering scores of all the noise features are uniformly very close to zero. Thus, if we use any arbitrarily small but prefixed value for the threshold $\alpha_0$, we have theoretical guarantees for \textit{perfectly screening out} all the noise coordinates.

It is important to have a small value of $\alpha_0$ to avoid screening out the informative features, which have non-negligible clustering scores~$S_j$.  Theorem~\ref{thm1} suggests that $\alpha_{0}^{\rm OR}=b\ln (p\vee n)/n$ is a reasonable choice.  Provided~$n$ is sufficiently large, an approach using the above choice of~$\alpha_0$ will not screen out the features identified as multi-cluster features by the \textit{population clustering procedure}, defined in Section~2.2 of \cite{radchenko2014consistent}.  The next result, which is a consequence of Theorem~\ref{thm1} above and Theorem~1 in \cite{radchenko2014consistent}, formalizes this point.  Note that \cite{radchenko2014consistent} demonstrate, through simulations and theoretical analysis, that the population procedure generally classifies multi-modal distributions as multi-cluster, provided the corresponding sub-populations are of reasonable size and have a moderate amount of separation.  To illustrate this fact, in \ref{appendix:normals} we provide a detailed summary of how the population procedure performs on a wide variety of bimodal Gaussian distributions.
\begin{corollary}
	Suppose that regularity conditions C1 and C2, stated in \ref{sec:appa}, are satisfied.  Let the cardinality of the set $\mathcal{I}_S$ be bounded above by a universal constant.  Suppose that the population clustering procedure identifies each feature in $\mathcal{I}_S$ as multi-cluster. Then, for all sufficiently small~$\tau>0$ there exist positive constants $c_1$, $c_2$, $b$ and~$\kappa$, whose choice does not depend on either~$n$ or~$p$, such that
	\begin{equation*}
	\Pr  ( \{j: S_{j}(\tau) >  \alpha_{0}^{\rm OR}  \} = \mathcal{I}_S ) \ge 1-c_1p^{-c_2},
	\end{equation*}
	as long as $p\le \exp(\kappa n)$.
	\label{cor1}
\end{corollary}
We view $\alpha_{0}^{\rm OR}$ as the \textit{oracle choice} of the threshold. However, it is difficult to evaluate it from the data,  primarily because the constant $b$ depends on the marginal densities $g_j$ of the noise features.   In the following section, we discuss several practical choices for the threshold parameter.

\section{Estimation of hyperparameters}
\label{sec3-est}

The COSCI procedure presented in Algorithm~1 requires only one tuning parameter, $\alpha_0$, as an input. The \textit{oracle threshold choice}~$\alpha_{0}^{\rm OR}$ is difficult to estimate from the data, as the marginal distributions of the features are typically unknown. In what follows, we present two approaches for estimating~$\alpha_0$ that are adaptive to the sample size~$n$, in the sense that a larger threshold is chosen for smaller sample sizes.
\subsection{Simulation based}
We generate data of varying sample sizes from several well-known unimodal distributions and use it to assess at which values of~$\alpha_0$ COSCI will detect no clusters, screen out the corresponding non-informative feature. The eight unimodal distributions considered in Table \ref{tabsim} are meant to represent the spectrum of the noise coordinates that are commonly encountered in real data applications. They include symmetric densities with the support equal to the entire \CG{real line $\mathbb{R}$}, densities with heavy tails and those with bounded support.

Table \ref{tabsim} in \ref{sec:appd} presents the results of this simulation exercise over $100$ repetitions. For example, when the noise coordinate is Gaussian, and the sample size is $n=500$, COSCI detects clusters in the majority of the~$100$ repetitions when $\alpha_0\le 0.05$. Thus, an appropriate threshold for this case should at least be greater than $0.05$. For larger sample sizes, COSCI detects no clusters with a relatively smaller $\alpha_0$. A general theme that emerges from this table is that when the underlying density is non-Gaussian with support over \CG{all of $\mathbb{R}$ or $\mathbb{R}_{+} = (0, \infty)$}, smaller thresholds seem to succeed at screening out the corresponding feature, when compared to the Gaussian case for the same sample size.  Similarly to the Gaussian case, densities with bounded support, such as the Beta and the Triangular distribution, require a larger threshold to succeed. Our practical recommendation is to assume an underlying Gaussian noise distribution and use $\alpha_0$ as the smallest threshold that detects no clusters given the sample size $n$. Let $\widehat{\alpha}_0$ be such a threshold. Then,
the selected feature set is $\mathcal{\widehat{I}}_{S}=\{j:S_j\ge\widehat{\alpha}_0\}$.
\subsection{Data driven}
\label{dd}
In this section, we use a data driven technique to estimate $\alpha_0$. We work under the large-scale multiple testing framework of \citet{efron2007size} and transform the problem of estimating $\alpha_0$ into a problem of feature selection using the merge sizes \CG{$S_1, \ldots , S_p$}. One can then read off the optimal $\alpha_0$ from the selected features as
$$
\widehat{\alpha}_{0}=\min_{j\in\mathcal{\widehat{I}}_{S}}S_j,
$$
where $\mathcal{\widehat{I}}_S$ holds the indices of the selected features.
Let $\psi_j=2S_j$ be the test statistic for testing the significance of cluster strength in feature $j$. Note that \CG{the} $\psi_j$'s have a mixture density $f$ given by $f(\psi) = \pi_0 f_{0}(\psi)+(1-\pi_0) f_{1}(\psi)$,
where $f_0$ is the \emph{theoretical null} distribution and $\pi_0$ is the null prior probability. The exact distributional form of $f_0$ is, however, unknown, primarily because we do not know the underlying distribution of the noise coordinates that generate the $S_j$'s. Nonetheless, Table~\ref{tabsim} ascertains that $f_0$ is \CG{right-skewed} on the support $[0,1]$, with  the mass concentrated around zero for large~$n$. We use the MLE method of \cite{efron2007size} to estimate the empirical null distribution from the observed $\psi_j$'s as a \CG{Beta} distribution and obtain the estimated false discovery rate (fdr)
\begin{equation}\label{fdr}
T_j = \widehat{\pi}_0\widehat{f}_0(\psi_j)/\widehat{f}(\psi_j)
\end{equation}
under the prescribed assumption that $\pi_0\ge 0.9$, i.e., \CG{at least} $90\%$ of the $p$ tests are null and $f_1(\psi)=0$ on $\mathcal{A}$, where $\mathcal{A}=\{\psi_{(1)}\le\cdots\le\psi_{([0.9p])}\}$. This scheme works reasonably well in both the simulations and the real data examples that we considered. For additional information about fdr see, e.g., Chap. 5 of \citet{efron2012large}. For estimating the mixture density $f$, we adopt Lindsey's method \citep{efron1996using,lindsey1974construction} that models the histogram bin counts using Poisson regression, treating the bin centers as covariates. Finally, to select the features we adopt the two-stage approach to signal screening recently introduced in \citet{tony2016optimal}, the details of which are relegated to \ref{appb}.

In the empirical analysis that follows, we estimate $\mathcal{\widehat{I}}_{S}$ and $\widehat{\alpha}_{0}$ using the method described above.

\section{Empirical analysis}
\label{sec4}

\subsection{Simulations}\label{sec:4.1} We perform several simulation experiments to gauge the feature screening performance of COSCI under two scenarios: (i) $p<n$ and (ii) $p>n$. For scenario (i), we implement two experiments and use the following four competing approaches of feature screening in clustering to compare the performance of our proposed method:
\begin{enumerate}
	\item Sparse $k$-means clustering (SpKM) and hierarchical clustering (SpHC) \citep{witten2012framework} --- we use the \textsf{R}-package {\texttt{sparcl}}.
	\item 	Sparse Alternate Sum (SAS) clustering \citep{arias2016simple} --- we use the \textsf{R} codes available at the author's site (see: \url{https://github.com/victorpu/SAS_Hill_Climb}).
	\item 	Influential Features PCA (IF-PCA) \cite{jin2016influential} --- we use the MATLAB codes available at the author's site (see \url{http://www.stat.cmu.edu/~jiashun/Research/software/HCClustering/}).
	\item Excess Mass method (Ex. Mass) \citep{chan2012using} for which unfortunately a software implementation  is not available in the public domain.
	Noting that the excess mass test and the dip test are equivalent in \CG{a} univariate setting \citep{cheng1998calibrating}, we implement a version of this method for coordinate-wise feature screening using Hartigan's Dip test \cite{hartigan1987estimation,hartigan1985dip}. For feature selection, we select those features for which the multiplicity adjusted \cite{benjamini1995controlling} \CG{$p$-values} from this test are at most 0.05.
\end{enumerate}
For each of the above methods, we are only interested in their feature selection capabilities and not on their clustering performance.

\begin{table}[t!]
	\centering
	\caption{False Negatives and False Positive rates for Simulation Experiment I. Here, $\mathcal{I}_S=\{1, \ldots,5\}$, $p=50$. The numbers in parenthesis are standard errors over $50$ repetitions.}
	\scalebox{0.8}{\begin{tabular}{crrrrrrr}
			\multicolumn{2}{c}{} & \multicolumn{2}{c}{$n = 200$} & \multicolumn{ 2}{c}{$n = 1000$} & \multicolumn{ 2}{c}{$n = 2500$}\\
			\toprule
			&  & Avg FN & Avg FP~~ & Avg FN~~ & Avg FP~~ & Avg FN~~ & Avg FP~~ \\
			\midrule
			\multicolumn{1}{c}{} & 0.05  & 0.06 (0.03) & 42.38 (0.22) & 0.10 (0.04) & 21.26 (0.53) & 0.14 (0.06) & 6.08 (0.33) \\
			\multicolumn{1}{c}{COSCI} & 0.08  & 0.22 (0.06) & 32.88 (0.35) & 0.24 (0.06) & 10.24 (0.41) & 0.34 (0.08) & 2.22 (0.23) \\
			\multicolumn{1}{c}{with} & 0.1   & 0.34 (0.07) & 26.96 (0.43) & 0.34 (0.07) & 7.14 (0.33) & 0.40 (0.08) & 1.20 (0.17) \\
			\multicolumn{1}{c}{$\alpha_0$} & 0.12  & 0.52 (0.08) & 22.12 (0.41) & 0.50 (0.1) & 5.16 (0.27) & 0.48 (0.09) & 0.82 (0.11) \\
			\multicolumn{1}{c}{fixed} & 0.15  & 0.74 (0.1) & 16.16 (0.48) & 0.76 (0.1) & 3.28 (0.19) & 0.64 (0.09) & 0.50 (0.10) \\
			\multicolumn{1}{c}{} & 0.2   & 1.12 (0.1) & 9.36 (0.40) & 1.04 (0.1) & 1.68 (0.15) & 0.92 (0.09) & 0.30 (0.08) \\
			\midrule
			\multicolumn{1}{c}{} &    Data driven & 1.8 (0.06) & 0.54 (0.13) & 0.88 (0.1) & 2.14 (0.17) & 0.36 (0.08) & 1.28 (0.10) \\
			\midrule
			\multicolumn{1}{c}{} &   SpKM & 2.96 (0.24) & 16.04 (2.28) & 3.26 (0.26) & 13.28 (2.27) & 3.9 (0.41)&  7.00 (4.19)\\
			\multicolumn{1}{c}{Other} &  SpHC & 3.34 (0.21) & 16.86 (1.15) & 2.08 (0.19) & 22.54 (0.77) & 1.6 (0.40) & 24.5 (2.02)\\
			\multicolumn{1}{c}{methods} &   SAS & 1.46 (0.14) & 19.66 (1.15) & 1.70 (0.18) & 19.52 (1.29) & 1.86 (0.20) & 17.04 (1.25) \\
			\multicolumn{1}{c}{} &  Ex. Mass & 2.3 (0.09) & 0.00 (0.00) & 2.00 (0.00) & 0.00 (0.00) & 2.00 (0.00) & 0.00 (0.00) \\
			\multicolumn{1}{c}{} &  IF-PCA & 0.32 (0.07) & 17.64 (0.93) & 0.10 (0.04) & 18.08 (0.86) & 0.10 (0.04) & 17.24 (0.99)\\
			\bottomrule
		\end{tabular}
		\label{simtab1}}%
\end{table}

For simulation Experiment~I, we consider features from a wide range of parametric distributions including correlated features and consider three different sample sizes from low to high.  Simulation Experiment~I  represents scenario (i); we fix $p=50$ and consider a design matrix ${X}_{n\times 50}$ with  $p_S=5$ and $p_N=45$. The $p_N$ noise coordinates are taken to be iid standard Gaussian, \CG{$\mathcal{N}(0,1)$}, while the $p_S$ signal coordinates are chosen as follows:
\begin{enumerate}
	\item $X_1 \sim 0.5\, \mathcal{B} (4,6) + 0.5 \,\mathcal{B} (7,3)$
	\item $X_2 \sim 0.5\, \mathcal{LN}(0.2,0.35)+0.5\, \mathcal{N}(4,0.5)$
	\item $X_3 \sim 0.5 \, \mathcal{L}(3,1.5) + 0.5 \mathcal{L}(5,1.5)$, \CG{where $\mathcal{L}$ refers to the Laplace or double exponential distribution,} and,
	\item $(X_4,X_5) \sim \sum_{i=1}^{4}\mathcal{N}({\mu}_i,\Sigma_i)/4$, where
	\begin{eqnarray*}
		{\mu}_1=(0,0), \quad {\mu}_2=(0,-4), \quad {\mu}_3=(4,0), \quad {\mu}_4={\mu}_3+{\mu}_2,\\
		\Sigma_1 =\Sigma_4=\begin{pmatrix}
			1 & -0.85\\
			-0.85 & 1
		\end{pmatrix}, \quad \Sigma_2 =\Sigma_3=\begin{pmatrix}
			1 & 0.85\\
			0.85 & 1
		\end{pmatrix}.
	\end{eqnarray*}
\end{enumerate}
In this setting, the signal coordinates are all bi-modal and with the exception of $X_1,X_3$, the separation between the adjacent medians is fairly large. For Experiment~II, we fix $p=100$ and consider a design matrix ${X}_{n\times 100}$ with  $p_S=6$ and $p_N=94$. We let half of the $p_N$ noise coordinates to be iid standard Gaussian and the other half to be iid \CG{Student $t$} random variables with 5 degrees of freedom. The $p_S$ signal coordinates are chosen as follows:
\begin{enumerate}
	\item $(X_1,\ldots,X_5)\sim $ as in Experiment~I
	\item $X_6\sim\, 0.3\,\mathcal{N}(-2.5,1) + 0.3\,\mathcal{N}(0,1) + 0.4\,\mathcal{N}(2.5,1)$ which is a non-symmetric, tri-modal  density.
\end{enumerate}

For each of the setups described above, columns of the data matrix ${X}$ are simulated for \CG{each $n \in \{200, 1000, 2500\}$}. We analyze two variants of COSCI: (i) COSCI with $\alpha_0$ fixed over a grid of six values, \CG{viz. $0.05, 0.08,0.1,0.12,0.15,0.2$,} and, (ii) COSCI with $\alpha_0$ estimated in a data driven fashion as discussed in Section \ref{dd}. For each method we calculated two statistics averaged over $50$ repetitions: False Negative (FN) --- the number of signal features incorrectly identified as noise; False Positive (FP) --- the number of noise features incorrectly identified as signal.

\begin{table}[b!]
	\centering
	\caption{False Negatives and False Positive rates for Simulation Experiment II. Here, $\mathcal{I}_S=\{1,\ldots ,6\}$, $p=100$. The numbers in parenthesis are standard errors over $50$ repetitions.}
	\scalebox{0.8}{\begin{tabular}{crrrrrrr}
			\multicolumn{2}{c}{} & \multicolumn{2}{c}{$n = 200$} & \multicolumn{2}{c}{$n = 1000$} & \multicolumn{2}{c}{$n = 2500$}\\ \toprule
			& & Avg FN & Avg FP~~ & Avg FN~~ & Avg FP~~ & Avg FN~~ & Avg FP~~ \\
			\midrule
			\multicolumn{1}{c}{} & 0.05  & 0.06 (0.03) & 84.76 (0.41) & 0.1 (0.04) & 34.62 (0.71) & 0.14 (0.06) & 8.2 (0.44) \\
			\multicolumn{1}{c}{COSCI} & 0.08  & 0.22 (0.06) & 61.9 (0.57) & 0.38 (0.08) & 15.18 (0.54) & 0.58 (0.11) & 2.64 (0.25) \\
			\multicolumn{1}{c}{with} & 0.1   & 0.38 (0.08) & 49.5 (0.64) & 0.52 (0.09) & 10.3 (0.40) & 0.76 (0.12) & 1.38 (0.18) \\
			\multicolumn{1}{c}{$\alpha_0$} & 0.12  & 0.6 (0.09) & 39.32 (0.62) & 0.72 (0.12) & 7.34 (0.30) & 0.96 (0.11) & 0.96 (0.13) \\
			\multicolumn{1}{c}{fixed} & 0.15  & 0.88 (0.1) & 28.16 (0.62) & 1.06 (0.13) & 4.86 (0.26) & 1.36 (0.12) & 0.58 (0.11) \\
			\multicolumn{1}{c}{} & 0.2   & 1.4 (0.11) & 15.9 (0.52) & 1.74 (0.12) & 2.46 (0.19) & 1.86 (0.10) & 0.34 (0.09) \\
			\midrule
			\multicolumn{1}{c}{} & Data driven & 1.66 (0.08) & 2.04 (0.28) & 0.86 (0.11) & 6.98 (0.24) & 0.18 (0.06) & 7.04 (0.27) \\
			\midrule
			\multicolumn{1}{c}{} & SpKM & 4.36 (0.25) & 29.04 (3.23) & 5.02 (0.26) & 18.42 (3.93) &   5.5 (0.50) &  14.00 (8.56) \\
			\multicolumn{1}{c}{Other} & SpHC & 3.52 (0.27) & 38.6 (3.05) & 1.54 (0.23) & 61.9 (3.18) & ---~~~~~~ &  ---~~~~~~ \\
			\multicolumn{1}{c}{methods} & SAS & 1.14 (0.15) & 58.48 (1.91) & 2.1 (0.28) & 41.66 (2.95) & 1.82 (0.27) & 43.9 (2.95) \\
			\multicolumn{1}{c}{} & Ex. Mass & 2.36 (0.10) & 0.00 (0.00) & 2.00 (0.00) & 0.00 (0.00) & 2.00 (0.00) & 0.00 (0.00) \\
			\multicolumn{1}{c}{} & IF-PCA & 1.96 (0.15) & 18.46 (2.24) & 0.3 (0.06) & 38.84 (0.61) & 0.14 (0.05) & 42.36 (0.23)\\
			\bottomrule
		\end{tabular}
		\label{simtab2}}
\end{table}

From Table \ref{simtab1}, it is evident that SpKM fails to detect \CG{at least three out of the five} signal features across the three sample sizes, while SpHC and SAS have relatively better FN performance. The SpHC algorithm faced \CG{scalability} issues (marked by dash in the table) for \CG{sample size $n=2500$. In contrast,} IF-PCA correctly identifies the five signal features. All of these four methods have high false positives. Variations in sample sizes do not seem to affect their performance in any significant manner. Ex. Mass consistently fails to identify multi-modality in $X_1$ and $X_3$ but has the best false positive performance across all the methods. This is not unexpected given that the noise features considered are uni-modal. For $\alpha_0$ small, COSCI identifies the five signal features (low FN rate) but also incorrectly includes many noise features as signals (high FP rate). As expected, this trend transitions into a high FN rate and low FP rate for larger $\alpha_0$'s. This is where the data driven approach to choose $\alpha_0$ is seen to be beneficial. For moderately large $n$, COSCI, coupled with the data driven approach, returns a FN rate comparable to IF-PCA but with a significantly smaller FP rate. The results in Table \ref{simtab2} reveal a similar picture. When both FN and FP rates are taken into consideration, COSCI delivers the best performance amongst all the competing methods for moderately large $n$. Even when $n=200$, COSCI returns a FN rate which is only slightly higher than IF-PCA and SAS but enjoys a far better FP rate.

The next two simulation experiments III and IV exemplify scenario (ii) where $p>n$. We consider noise features which include non-symmetric distributions like \CG{the} Exponential with rate parameter 1, \CG{$\mathcal{E}(1)$}, and heavy-tailed distributions like \CG{the} standard Cauchy. This presents an interesting setting especially for methods like IF-PCA that rely on statistical comparison with a fixed parametric distribution to determine feature importance for clustering. In experiment III, we have $p=5000$ and $p_S=7$ with
\begin{enumerate}
	\item $(X_1,\ldots,X_6)\sim $ as in Experiment I
	\item $X_7\sim 0.5\, \mathcal{N}(-1.1,1) + 0.5\, \mathcal{N}(1.1 ,1)$.
\end{enumerate}

Note that $X_7$ is a bi-modal Gaussian mixture but with a relatively small separation between the modes. The $p_N$ noise features consists of  approximately $40\%$ iid \CG{$\mathcal{E}(1)$} noise \CG{along with} $30\%$ iid standard Gaussian and $30\%$ iid $t_5$ noise. For Experiment~IV, we consider $p=\mbox{25,000}$ and add two more signal features so that $p_S=9$ with
\begin{enumerate}
	\item $(X_1,\ldots,X_7)\sim $ as in Experiment III
	\item $X_8\sim 0.3 \mathcal{L}(-3,1) + 0.35 \mathcal{L}(0,1) + 0.35 \mathcal{L}(3,1)$.
	\item $X_9\sim 0.3\, \mathcal{B}(8,2) + 0.35\,\mathcal{B}(5,5) + 0.35\,\mathcal{B}(2,8)$.
\end{enumerate}

Here $X_8,X_9$, once again, represent features with a relatively small separation between the modes and, thus, are particularly difficult examples for modality detection. The $p_N$ noise features here include approximately $28\%$ iid standard Cauchy noise \CG{along with} $24\%$ iid standard Gaussian, $24\%$ iid {t}$_5$ and $24\%$ iid $\mathcal{E}(1)$ noises. We keep all the other design parameters of Experiment~IV identical to Experiment~III.  However, we do not include the performance of SpKM, SpHC and SAS in our analysis of Experiment~IV as these algorithms are computationally very demanding and often exhibited convergence issues  in this regime.

\begin{table}[b!]
	\centering
	\caption{False Negatives and False Positive rates for Simulation Experiment III. Here, $\mathcal{I}_S=\{1,\ldots ,7\}$, $p=5000$. The numbers in parenthesis are standard errors over $50$ repetitions.}
	
	\medskip
	\scalebox{0.75}{\begin{tabular}{crrrrrrr}
			\multicolumn{2}{c}{} & \multicolumn{2}{c}{$n = 200$} & \multicolumn{2}{c}{$n = 1000$} & \multicolumn{2}{c}{$n = 2500$}\\ \toprule
			& & Avg FN & Avg FP~~ & Avg FN~~ & Avg FP~~ & Avg FN~~ & Avg FP~~ \\
			\midrule
			\multicolumn{1}{c}{} & 0.05  & 0.12 (0.05) & 4341.6 (3.13) & 0.82 (0.08) & 1,355.3 (3.84) & 1.06 (0.07) & 274.62 (2.56) \\
			\multicolumn{1}{c}{COSCI} &  0.08  & 0.48 (0.08) & 2979.62 (5.06) & 1.26 (0.09) & 548.8 (3.11) & 1.58 (0.11) & 82.86 (1.30) \\
			\multicolumn{1}{c}{with} &  0.1   & 0.84 (0.11) & 2279.44 (4.37) & 1.44 (0.1) & 349.68 (2.04) & 1.76 (0.12) & 47.5 (1.08) \\
			\multicolumn{1}{c}{$\alpha_0$} &  0.12  & 1.22 (0.12) & 1764.22 (5.09) & 1.68 (0.12) & 242.7 (1.75) & 1.96 (0.11) & 30.8 (0.83) \\
			\multicolumn{1}{c}{fixed} &  0.15  & 1.56 (0.13) & 1215.96 (4.86) & 2.04 (0.13) & 151.74 (1.21) & 2.36 (0.12) & 17.64 (0.62) \\
			\multicolumn{1}{c}{} &  0.2   & 2.22 (0.12) & 636.62 (3.68) & 2.72 (0.13) & 69.58 (0.96) & 2.86 (0.10) & 7.74 (0.44) \\
			\midrule
			\multicolumn{1}{c}{} & Data driven & 2.68 (0.12) & 297.14 (6.50) & 1.32 (0.10) & 498.76 (2.86) & 0.92 (0.07) & 516.28 (4.19) \\
			\midrule
			\multicolumn{1}{c}{Other} & Ex. Mass & 4.86 (0.14) & 0.00 (0.00) & 4.00 (0.00) & 0.00 (0.00) & 4.00 (0.00) & 0.00 (0.00) \\
			\multicolumn{1}{c}{methods} & IF-PCA & 5.68 (0.09) & 1802.4 (4.29) & 6.00 (0.00) & 1943.6 (1.86) & 6.00 (0.00) & 1970.00 (0.88)\\
			\bottomrule
		\end{tabular}
		\label{simtab3}}%
\end{table}

\begin{table}[t!]
	\centering
	\caption{False Negatives and False Positive rates for Simulation Experiment IV. Here,  $\mathcal{I}_S=\{1,\ldots,9\}$, $p=\mbox{25,000}$. The numbers in parenthesis are standard errors over $10$ repetitions.}\scalebox{0.75}{\begin{tabular}{crrrrrrr}
			\multicolumn{2}{c}{} & \multicolumn{2}{c}{$n = 200$} & \multicolumn{2}{c}{$n = 1000$} & \multicolumn{2}{c}{$n = 2500$}\\ \toprule
			& & \multicolumn{1}{c}{Avg FN}~~ & \multicolumn{1}{c}{Avg FP}~~ & Avg FN~~ & Avg FP~~ & Avg FN~~ & Avg FP~~ \\
			\midrule
			\multicolumn{1}{c}{} &  0.05  & \multicolumn{1}{c}{0.12 (0.05)} & \multicolumn{1}{c}{18587.2 (8.03)} & 0.6 (0.22) & 5310.00 (22.81) & 1.00 (0.1) & 1101.67 (11.5) \\
			\multicolumn{1}{c}{COSCI} &  0.08  & \multicolumn{1}{c}{0.48 (0.08)} & \multicolumn{1}{c}{12176.00 (10.6)} & 1.30 (0.3) & 2170.00 (11.71) & 1.47 (0.192) & 339.2 (4.82) \\
			\multicolumn{1}{c}{with} &  0.1   & \multicolumn{1}{c}{0.84 (0.11)} & \multicolumn{1}{c}{9221.48 (10.01)} & 1.50 (0.27) & 1397.5 (9.96) & 1.6 (0.19) & 196 (2.68) \\
			\multicolumn{1}{c}{$\alpha_0$} &  0.12  & \multicolumn{1}{c}{1.22 (0.12)} & \multicolumn{1}{c}{7088.34 (9.18)} & 1.70 (0.26) & 976.5 (8.16) & 1.73 (0.18) & 126 (2.99) \\
			\multicolumn{1}{c}{fixed} &  0.15  & \multicolumn{1}{c}{1.60 (0.13)} & \multicolumn{1}{c}{4862.02 (8.61)} & 2.10 (0.28) & 616.40 (7.22) & 2.00 (0.22) & 72.87 (2.41) \\
			\multicolumn{1}{c}{} &  0.2   & \multicolumn{1}{c}{2.42 (0.12)} & \multicolumn{1}{c}{2537.76 (6.92)} & 2.60 (0.27) & 279.5 (3.27) & 2.8 (0.22) & 29.07 (1.65)\\
			\midrule
			\multicolumn{1}{c}{} & Data driven & \multicolumn{1}{c}{2.84 (0.14)} & \multicolumn{1}{c}{1687.34 (17.77)} & 1.40 (0.30) & 1928.3 (18.72) & 1.00 (0.1) & 1324.8 (19.27) \\
			\midrule
			\multicolumn{1}{c}{Other} & Ex. Mass & \multicolumn{1}{c}{7.16 (0.13)} & \multicolumn{1}{c}{0.00 (0.00)} & 6.00 (0.00) & 0.00 (0.00) & 5.2 (0.11) & 0.00 (0.00) \\
			\multicolumn{1}{c}{methods} & IF-PCA & 9.00 (0.00) & 4061.2 (17.71) & 9.00 (0.00) & 5286.5 (17.64) & 9.00 (0.00) & 5563.3 (12.97)\\
			\bottomrule
		\end{tabular}
		\label{simtab4}}
\end{table}

From Tables \ref{simtab3} and \ref{simtab4}, a drop in the FN performance of both IF-PCA and Ex. Mass is conspicuous. For IF-PCA, the non-Gaussian noises are identified as signals which are ultimately selected in favor of the true signals under the HC functional. \CG{In contrast,} Ex. Mass continues to conclude \CG{that} some of the difficult multi-modal signals like $X_1,X_3,X_7,X_8$ and $X_9$ \CG{are} unimodal. COSCI, with the data driven approach, once again returns the best performance across all the different sample size regimes considered in these two experiments.

\subsection{Real data examples}
\label{sec:realdata}

We test the performance of COSCI on a number of real data examples. In all these datasets, the number of clusters / subpopulations is known \CG{a priori}. However, unlike the simulation study, we do not know the ``true'' feature set for these datasets and thus an estimate of screening performance based on False Negatives or False Positives is impossible. Instead, we use the following scheme to gauge the performance of COSCI. On each of the datasets considered in this section, we overlay COSCI with $k$-means (KM), Sparse $k$-means (SpKM) and IF-PCA. In other words, we allow COSCI to screen and select the best features and thereafter, we run classical $k$-means, Sparse $k$-means and IF-PCA on the selected features. While using IF-PCA on the COSCI screened features, we only use the clustering component of IF-PCA and not \CG{its} feature selection step. The classification error rates (CER) so obtained from the above \CG{three} schemes are then compared to the CER of the competing methods, \CG{i.e.,} $k$-means, Sparse $k$-means and IF-PCA, all without any COSCI screening. For $k$-means and IF-PCA, we report the average CER over 30 independent replications and the associated standard error whenever they are bigger than $0.0005$. Next, we describe in details the application of the aforementioned methods on three real \CG{datasets}. In \ref{sec:appc1} the classification results on eleven other datasets are also demonstrated.

\medskip
\noindent\textit{Multi-tissue data:} This is a microarray data on different mammalian tissue types. The data \CG{were} produced by \citet{su2002large} and \CG{they hold} gene expression from human and mouse samples across a diverse array of tissues, organs and cell lines. There are $n=102$ samples and $p=5565$ genes  in \CG{these} data. The tissue types have four categories that are known and the goal is to identify the sub-populations that correspond to the four tissue types. \CG{These} data can be publicly sourced from the \textsf{R}-package {\texttt{FABIA}} \citep{fabia}.
\begin{figure}[!h]
	\centering
	\includegraphics[height=20pc,width=32pc]{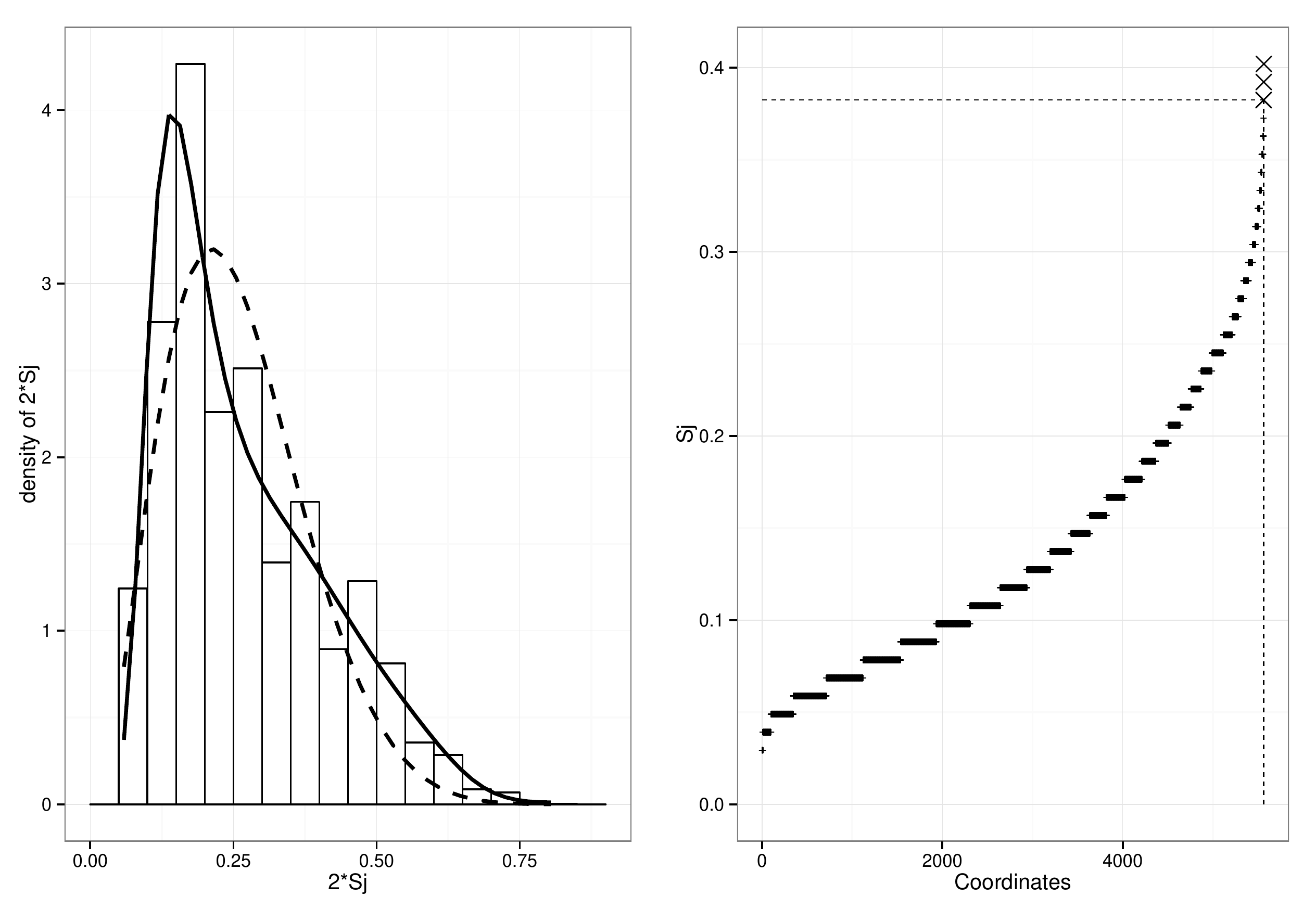}
	\caption{\textsc{Multi-tissue Data.} Left: $\widehat{\pi}_{0}\widehat{f_{0}}(\psi_j)$ in black dashed line and $\widehat{f}(\psi_j)$ in black solid line for \CG{all $j \in\{1,\ldots,5565\}$}. Right: Distribution of $S_j$. The \CG{four} selected features are marked as $\times$. The dashed horizontal line is $\widehat{\alpha}_0=0.382$.}
	\label{multif}
\end{figure}
In \CG{Figure}~\ref{multif}, we plot the distribution of $\psi_j$ (left) \CG{along with} the rank ordering of the 5565 features with respect to the scores $S_j$ (right). Using the data driven selection procedure, COSCI selects the top \CG{four} features and returns an estimate of $\alpha_0$ almost close to $0.4$ which agrees well with the threshold choice prescribed in Table \ref{tabsim} for sample size 100 and  Gaussian noise coordinates.  CERs for the aforementioned classification methods are reported in Table~\ref{realcer1}.

\medskip
\noindent\textit{Cardio data:} This data has $n=63$ subjects of which 19 are cardiovascular patients and the rest are healthy controls. The genetic expression of each subject has been recorded for $p = \mbox{20,426}$ genes. The goal is to classify the subjects as healthy or cardiovascular based on expression levels of the $p$ genes. The dataset is publicly available on Brad Efron's webpage.
In \CG{Figure~}\ref{cardiof} (\ref{sec:appc2}), we plot the distribution of $\psi_j$ (left) \CG{along with} the rank ordering of the 20,426 features with respect to the scores $S_j$ (right). Using the data driven selection procedure, COSCI selects the top 33 features and results in an estimate of $\alpha_0$ equal to $0.428$. Due to the small sample size of this data, the data driven procedure returns an unrealistic estimate of null proportion $\widehat{\pi}_0=0.99$ when the empirical null distribution is estimated on $\mathcal{A}=\{\psi_{(1)}\le\cdots\le\psi_{([0.9p])}\}$. We comment more on this observation after introducing the results in Table~\ref{realcer1}.

\medskip
\noindent\textit{RNASeq data:}\label{rnasec}
The RNASeq data discussed in Section \ref{sec1} is an example where both $n$ and $p$ are large. For such large values of $n$, one would expect an asymptotic regime to kick in and hope to see the distribution of $\psi_j$'s concentrated around a small value. In \CG{Figure}~\ref{rnaf} (Appendix \ref{sec:appc2}) (left), we see that approximately $70\%$ of the coordinates have $\psi_j\le 0.1$. Moreover, the distribution of $S_j$ (figure \ref{rnaf} right) has an explicit ``elbow'' at $0.1$, suggesting that we could take the cut-off on $\alpha_0$ to be 0.1 and select those features with $S_j \ge 0.1$. This is indeed one of the ways that we can approach the problem given this data and that would have left us with \CG{approximately} $ 2800$ features. We, however, continue with the data driven approach but take $\mathcal{A}=\{\psi_{(1)}\le\cdots\le\psi_{([0.7p])}\}$. This leads to an estimated null proportion $\widehat{\pi}_0=0.70$ with the top 2304 features as the selected ones. Estimated $\widehat{\alpha}_0$ is 0.188 which (i) is not far from the ``elbow'' in the distribution of $S_j$ and (ii) once again agrees well with the theoretical choice prescribed under Gaussian noises in Table \ref{tabsim}.

\begin{table}[b!]
		\centering
	\caption{This table has \CG{three} parts. In the first part of the table, we provide some preliminary information on the \CG{three} datasets. The second part of the table has CER for the \CG{three} competing methods. For IF-PCA, we also report the number of selected features after the `/' symbol. In the last part of the table, CER for COSCI + ``clustering method'' is reported \CG{along with} the number of selected features. The numbers in parenthesis are standard errors over 30 repetitions of $k$-means and IF-PCA. The standard errors are only reported when they exceed 0.0005. The SAS algorithm faced \CG{scalability} issues (marked by dash) for RNASeq data}
	
	\medskip
	\scalebox{0.95}{\begin{tabular}{cccccc}
			\multicolumn{2}{c}{} & \multicolumn{1}{c}{Multi-tissue} & \multicolumn{1}{c}{Cardio} & \multicolumn{1}{c}{RNASeq} \\
			\toprule
			\multicolumn{1}{c}{\;\;\;\;\;\;\;\;\;\;\;\; \;\; Prelim.} & $n$ & 102 & 63 & 2730 \\
			\multicolumn{1}{c}{\;\;\;\;\;\;\;\;\;\;\;\; \;\;info.} & $p$ & 5565 & 20426 & 8716 \\
			\multicolumn{1}{c}{} & \# clusters & 4 & 2 &  19\\
			\midrule
			\multicolumn{1}{c}{} & SpKM & 0.698 & 0.492 &  0.157\\
			\multicolumn{1}{c}{\;\;\;\;\;\;\;\;\;\;\;\; \;\;Competing} & KM & 0.721 (0.002) & 0.493 & 0.157 \\
			\multicolumn{1}{c}{\;\;\;\;\;\;\;\;\;\;\;\; \;\;methods} & SAS & 0.554 / 1307 & 0.492 / 1071 & --- \\
			\multicolumn{1}{c}{} & Ex. Mass & no selection & no selection & 0.157 / 6814 \\
			\multicolumn{1}{c}{} & IF-PCA & no selection & \textit{0.486 / 523} &  0.213 (0.002) / 4358\\
			\midrule
			\multicolumn{1}{c}{} & SpKM & 0.396 / 4 & 0.505 / 33 & \textit{0.152 / 2304} \\
			\multicolumn{1}{c}{\;\;\;\;\;\;\;\;\;\;\;\; \;\;COSCI +} & KM & 0.397 / 4 & 0.505 / 33 &  0.153 / 2304\\
			\multicolumn{1}{c}{} & IF-PCA & \textit{0.381 / 4} & 0.505 / 33 &  0.156 / 2304\\
			\bottomrule
		\end{tabular}
		\label{realcer1}}
\end{table}

We make several comments on the CER's reported in Table~\ref{realcer1}. The general theme of those results is compelling. These examples suggest that feature screening by COSCI can potentially lead to improvements in clustering error rates even when the underlying clustering algorithm is vanilla $k$-means. On the Multi-tissue data, IF-PCA and Ex. Mass fail to select any feature whereas the \CG{four} COSCI screened features overlaid with IF-PCA clustering clearly demonstrates substantial improvement over the competing error rates. A similar observation follows when COSCI is overlaid with Sparse $k$-means and $k$-means. On the RNASeq data, $k$-means performs well but in this example too, COSCI screening leads to an overall improvement in the error rate with much fewer features. On the Cardio data, performance enhancement is not observed primarily due to the small sample size of this dataset. For such low values of $n$, the added dispersion in the empirical null often masks relatively weaker signals from being identified. In these cases, the empirical null distribution may be estimated on a slightly smaller set $\mathcal{A}=\{\psi_{(1)}\le\cdots\le\psi_{([0.85p])}\}$ to improve clustering accuracy at the expense of a few more features. For example, on the Cardio data with $\mathcal{A}$ as prescribed above, COSCI selects only 129 features but returns a comparable CER of 0.486 using $k$-means.

\subsection{Beyond marginal screening: Two-way interactions}
\label{sec:combinations}

In this section we present an extension of COSCI that can successfully identify pairs of features that hold cluster information jointly but are un-informative marginally.
We expect that these pairs of features will reveal their inherent cluster strength through a suitable linear combination of the form ${X}^{i,j}u$ where ${X}^{i,j}_{n\times 2}$ is the feature pair using the $(i,j)$ feature in ${X}$ and $u\in \mathbb{R}^2$ with $||u||_2=1$. To determine the optimal $u$ for each feature pair $(i,j)$, we use a grid search in $\mathbb{R}^2$ as follows.
\begin{enumerate}
	\item[1.]Generate a uniform grid of $m$ points $u_k$ on the unit circle in $\mathbb{R}^2$.
	\item[2.] For the feature pair $(i,j)$, define ${Y}_{n\times m}=\{{X}^{i,j}\,u_k:1\le k\le m\}$ and use Algorithm~1 to get $S_{(i,j)}(u_1),\ldots,S_{(i,j)}(u_m)$ for the $m$ features in ${Y}$.
	\item[3.]Obtain the feature score for the pair $(i,j)$ as $S_{(i,j)} = \max_{1\le k\le m}S_{(i,j)}(u_k)$ and the optimal $u$ as  $u^{*}_{(i,j)}=\arg\max_{1\le k\le m}$ $S_{(i,j)}(u_k)$.
\end{enumerate}
We repeat this procedure for all the \CG{$p(p-1)/2$} feature pairs and choose
\begin{equation*}
\mathcal{\widehat{I}}_S=\text{unique}\, ( \{i:S_i\ge\alpha_0 \}\cup \{(i,j):S_{(i,j)}\ge\alpha_{0} \} ).
\end{equation*}
Often a component of $u^{*}_{(i,j)}$ will be small indicating that one of the contributing features in the pair $(i,j)$ dominates the other in terms of cluster strength. In these scenarios, we may run into the problem of including a lot of redundant pairs in $\mathcal{\widehat{I}}_S$ due to the strong effect of only one of the features. Such issues are easily resolved by using  a naive thresholding rule on $u^{*}_{(i,j)}$ to select the features and modifying $\mathcal{\widehat{I}}_S$ accordingly:
\begin{eqnarray*}
	\mathcal{\widehat{I}}_{S}^{\rm thr}&=&\begin{cases}
		 \{(i,j):S_{(i,j)}\ge\alpha_{0}\text{ and }\, \max_{\ell=1,2}|u^{*}_{(i,j)}(\ell)|<0.95 \},\\
		\{i:S_{(i,j)}\ge\alpha_{0}\text{ and }\, |u^{*}_{(i,j)}(1)|\ge0.95 \},\\
		\{j:S_{(i,j)}\ge\alpha_{0}\text{ and }\, |u^{*}_{(i,j)}(2)|\ge0.95 \},
	\end{cases}\\
	\mathcal{\widehat{I}}_S^{\sf \,mod}&=&\text{unique}\, ( \{i:S_i\ge\alpha_0 \}\cup\mathcal{\widehat{I}}_{\,S}^{\rm thr} ).
\end{eqnarray*}
To demonstrate the effectiveness of this approach, we consider a simple simulation setting (Experiment~V) with $m=20$, $p_S=4$, $p_N=21$, and $n=2000$. We let the $p_N$ noise features be iid Standard Normal, \CG{$\mathcal{N}(0,1)$}. For the signal coordinates, we take
\begin{enumerate}
	\item[1.] $(X_1,X_2) \sim \sum_{i=1}^{2}\mathcal{N}({\mu}_i,\Sigma)/2$
	\item[2.] $X_3 \sim 0.5 \, \mathcal{B} (4,6) + 0.5 \,\mathcal{B}(7,3)$
	\item[3.] $X_4 \sim 0.5 \, \mathcal{LN}(0.2,0.35) + 0.5 \,\mathcal{N}(4,0.5)$, where
	\begin{eqnarray*}
		{\mu}_1=(0.9,-0.9),\quad {\mu}_2=-{\mu}_1,\quad
		\Sigma =\begin{pmatrix}
			1 & 0.9\\
			0.9 & 1
		\end{pmatrix}
	\end{eqnarray*}
\end{enumerate}
In this setting features $X_3$ and $X_4$ are bi-modal but $(X_1,X_2)$ are only jointly bi-modal. Note that the effective dimensionality of the data in this example is \CG{$p_S+p_N+m (p_S+p_N)(p_S+p_N-1)/2=6025$}. In \CG{Table}~\ref{tab:2way}, we report the \textit{False Positive} and \textit{False Negative} proportions produced by applying the aforementioned extension of our proposed COSCI algorithm.  It successfully identifies all the signal coordinates. As expected, higher values of the threshold parameter improve the FP rate, and the benefit of using the data driven approach to select the features is evident. Once again, the prescribed theoretical choice (Table \ref{tabsim}), for the Gaussian noises (with $n = 2000$) agrees with these results. Table~\ref{tab:2way_cn} of the Appendix shows that the performance of the COSCI procedure is not affected even when the noises are significantly correlated with the features.

\begin{table}[t!]
	\centering
	\caption{False Negatives and False Positive rates for Simulation Experiment V. Here, $\mathcal{I}_S=\{1,2,3,4\}$, $p=25$. The numbers in parenthesis are standard errors over $10$ repetitions.}
	
	\medskip
	\scalebox{0.9}{\begin{tabular}{ccccccc}

			\multicolumn{1}{c}{} &       & \multicolumn{4}{c}{$n = 2000$ / $p = 25$} & \\
			\toprule
			&       & \multicolumn{2}{c}{Avg FN} & \multicolumn{2}{c}{Avg FP} \\
			\midrule
			\multicolumn{1}{c}{} & 0.05  & 0.00 &(0.00) & 21.00 &(0.00) \\
			\multicolumn{1}{c}{COSCI} & 0.08  & 0.00 &(0.00) & 21.00 &(0.00) \\
			\multicolumn{1}{c}{with} & 0.1   & 0.00 &(0.00) & 20.9 &(0.10) \\
			\multicolumn{1}{c}{$\alpha_0$} & 0.12  & 0.00 &(0.00) & 20.3 &(0.26) \\
			\multicolumn{1}{c}{fixed} & 0.15  & 0.00 &(0.00) & 18.8 &(0.39) \\
			\multicolumn{1}{c}{} & 0.2   & 0.00 &(0.00) & 13.7 &(0.68) \\
			\multicolumn{1}{c}{} & 0.25   & 0.00 &(0.00) & 3.8& (0.78) \\
			\midrule
			& Data driven & 0.00 (0.00)& 3.1 (0.23)\\
			\bottomrule
		\end{tabular}
		\label{tab:2way}}
\end{table}

\section{Discussion}
\label{sec5}

We propose COSCI, a novel feature screening method for large scale cluster analysis problems that are characterized by both large sample sizes and high dimensionality of the observations. COSCI efficiently ranks the candidate features in a \CG{nonparametric} fashion and, under mild regularity conditions, is robust to the distributional form of the true noise coordinates. We establish theoretical results supporting ideal feature screening properties of our proposed procedure and provide a data driven approach for selecting the screening threshold parameter. Extensive simulation experiments and real data studies demonstrate encouraging performance of our proposed approach.

An interesting topic for future research is extending our marginal screening method by means of utilizing multivariate objective criteria, which are more potent in detecting multivariate cluster information among marginally unimodal features.  Preliminary analysis of the corresponding $\ell_2$ fusion penalty based criterion, which, unlike the~$\ell_1$ based approach used in this paper, is non-separable across dimensions, suggests that this criterion can provide a way to move beyond marginal screening.

\appendix

\section{}
\subsection{Regularity conditions and proofs}
\label{sec:appa}

We write $g_j$ for the marginal density of the standardized \CG{$j$th} feature, $(X_j-\mathrm{E}X_j)/\mathrm{SD}(X_j)$, and let $q_j$ denote the corresponding quantile function.  Given feature~$j$, an interval $(\ell, r)$ and a point $a\in(\ell, r)$, we define the corresponding population criterion function as $G^j_{\ell, r}(a)=\mu^j_{a,R}-\mu^j_{L,a}$, where $\mu^j_{\ell, r}$ is the conditional mean on $(\ell, r)$ under the population distribution of the standardized \CG{$j$th} feature.  Let $P_{nj}$ denote the empirical measure associated with the observations in the \CG{$j$th} feature and let $P_j$ be the corresponding population distribution. We use the term population cluster to refer to all intervals that appear along the path of the corresponding univariate population splitting procedure, as defined in \cite{radchenko2014consistent}.

We now formally state the regularity conditions needed in Theorem~\ref{thm1}.    We suppose that there exist positive universal constants $C_{\tau}$ and $c_{\tau}$, which depend only on~$\tau$, and a positive universal~$C$, such that for all $j\in\mathcal{I}_N$ and $\tau\in(0,1)$:
\begin{enumerate}
	\item[C1:] The densities $g_j$ are differentiable and unimodal.  Also,
	\begin{eqnarray}
	\label{c2} \max_{j\in\mathcal{I}_N}\;\|g_j\|_{\infty}+\|g'_j\|_{\infty}&<&C,\\
	\label{c3} \max_{j\in\mathcal{I}_N}\;|q_j(\tau)|+|q_j(1-\tau)|&<&C_{\tau},\\
	\label{c4} \min_{j\in\mathcal{I}_N}\;g_j\{q_j(\tau)\} \wedge g_j\{q_j(1-\tau)\}&>&c_{\tau}.
	\end{eqnarray}
	\item[C2:] For each population cluster $(\ell, r)$ of the \CG{$j$th} feature, satisfying inequality $\int_L^R g_j(x)dx \ge 0.49$, and each $t\in \{L,R \}\cap \arg\max G_{\ell, r}^{j} \cap  [q_j(\tau),q_j(1-\tau) ]$, we have $ |G^{j \, \prime}_{\ell, r}(t) |>c_{\tau}$.
\end{enumerate}
Condition C1 ensures that we have uniform control over the noise densities.  Condition C2 is an appropriate adaptation of a standard regularity condition in M-estimation.  As we mentioned earlier, C1 and C2 are satisfied for location-scale families of unimodal differentiable densities with a finite first moment.  The assumption that~$g_j$ are differentiable can be slightly relaxed.  However, we prefer to keep this assumption, as it simplifies the presentation of the results.

\subsubsection{Proof of Theorem~\ref{thm1} and Corollary~\ref{cor1}}

We will use ``$\lesssim$'' to mean that inequality ``$\le$'' holds when the \CG{right-hand} side is multiplied by a positive constant, which is chosen independently from the parameters $p$, $n$, $j$ and $c_3$ (note that the constant is allowed to depend on~$\tau$ and $\epsilon_1$).  To simplify the exposition, we will write~$p$ for $p_N$, and have index $j$ always correspond to the (noise) coordinates in $\mathcal{I}_N$.

Note that the COSCI approach is invariant to linear transformations.  We will assume throughout the proof, without loss of generality, that the underlying distribution of each feature has mean \CG{$0$} and variance \CG{$1$}.  As a consequence of standardization, we have
\begin{equation}
\label{c1} \max_{j\in\mathcal{I}_N}\;\mathrm{E}|X_j|\le 1.
\end{equation}

Let $L_j$ and $R_j$ denote the smallest and the largest value, respectively, of the (univariate) cluster formed in the~\CG{$j$th} coordinate after the merge corresponding to~$S_j(\tau)$.  For the same merge, let $a_j$ be the midpoint between the closest representatives of the two sub-clusters.  We define the empirical criterion functions as $\widehat{G}^j_{\ell, r}(a)=\widehat{\mu}^j_{a,R}-\widehat{\mu}^j_{L,a}$, where $\widehat{\mu}_{\ell,r}$ denotes the average of the observations on the \CG{$j$th} feature that fall in $[\ell,r]$.
We need the following lemma, which is proved in~\ref{lemsec}.
\begin{lemma}
	\label{lem1}
	For every positive $\epsilon_1$ and $\tau$, there exist positive constants $c_0$, $c_3$, $c_4$, $c_5$ and~$\kappa$, such that, as long as $p\le \exp(\kappa n)$, the following inequalities simultaneously hold for all $j$ with probability bounded below by $1-c_4p^{-c_5}$:
	\begin{equation}
	\label{ap2}
| \{\widehat{G}_{L_j,R_j}^{j}(R_j)-\widehat{G}_{L_j,R_j}^{j}(a_j) \}- \{G_{L_j,R_j}^{j}(R_j)-G_{L_j,R_j}^{j}(a_j) \} |
\le \epsilon_1 (R_j-a_j)+c_3\dfrac{\ln (p\vee n)}{n},
	\end{equation}
	\begin{equation}
	\label{ap3}
| \{\widehat{G}_{L_j,R_j}^{j}(L_j)-\widehat{G}_{L_j,R_j}^{j}(a_j) \}- \{G_{L_j,R_j}^{j}(L_j)-G_{L_j,R_j}^{j}(a_j) \} |\\
\le \epsilon_1 (a_j-L_j)+c_3\dfrac{\ln (p\vee n)}{n} \, ,
	\end{equation}
	\begin{equation}
	\label{ap5}
	 |P_{nj}(R_j-a_j)-P_j(R_j-a_j) |\le \epsilon_1 (R_j-a_j)+c_3\dfrac{\ln (p\vee n)}{n},
	\end{equation}
	\begin{equation}
	\label{ap6}
	 |P_{nj}(a_j-L_j)-P_j(a_j-L_j) |\le \epsilon_1 (a_j-L_j)+c_3\dfrac{\ln (p\vee n)}{n} \, ,
	\end{equation}
	\begin{equation}
	\label{ap4}
\{{G}^{j}_{L_j,R_j}(L_j)-{G}^{j}_{L_j,R_j}(a_j) \}\vee \{{G}^{j}_{L_j,R_j}(R_j)-{G}^{j}_{L_j,R_j}(a_j)\}
\ge c_{0}(R_j-a_j)\wedge(a_j-L_j).
	\end{equation}
\end{lemma}

\medskip
For the remainder of the proof we restrict our attention to the set on which inequalities \eqref{ap2}--\eqref{ap4} are valid.  It follows directly from Proposition~2 in \cite{radchenko2014consistent} that $a_j\in\arg\max \widehat{G}^{j}_{L_j,R_j}$.
Taking $\epsilon_1=c_{0}/2$ in Lemma~\ref{lem1}, we derive
\begin{eqnarray}
0&\ge&  \{\widehat{G}^{j}_{L_j,R_j}(L)-\widehat{G}^{j}_{L_j,R_j}(a_j) \}\vee \{\widehat{G}^{j}_{L_j,R_j}(R_j)-\widehat{G}^{j}_{L_j,R_j}(a_j) \}\nonumber\\
&\ge& c_{0}(R_j-a_j)\wedge(a_j-L_j)-\epsilon_1(R_j-a_j)\wedge(a_j-L_j)+c_3\dfrac{\ln (p\vee n)}{n}\nonumber\\
&\ge& (c_{0}/2)(R_j-a_j)\wedge(a_j-L_j)-c_3\dfrac{\ln (p\vee n)}{n}\nonumber
\end{eqnarray}
Hence, for all~$j$, we have
\begin{equation*}
(R_j-a_j)\wedge(a_j-L_j)\le 2c_{0}^{-1}c_3\dfrac{\ln (p\vee n)}{n}.
\end{equation*}
Because the densities are uniformly bounded, the left-hand side can be replaced by $P_j(R_j-a_j)\wedge P_j(a_j-L_j)$ at the cost of an additional universal multiplicative factor on the right-hand side.  To complete the proof, it is only left to replace each $P_j$ with the corresponding empirical probability, $P_{nj}$.  This replacement is justified (again, at the cost of an additional universal multiplicative factor) by applying inequalities~\eqref{ap5} and~\eqref{ap6}.

\subsubsection{Proof of Lemma \ref{lem1}}\label{lemsec}
We will focus on inequality~\eqref{ap2} and, for concreteness, suppose that
$a_j\in [(L_j+R_j)/2, R_j ]$.  The rest of the cases can be handled using analogous arguments.

Applying Theorem~2.14.9 in \cite{van1996weak}, together with the union bound, we note that, for each positive~$\epsilon$, we have
\begin{equation*}
\Pr \Bigl\{\max_j\sup_{\ell<r}|P_{nj}(\ell,r)-P_j(\ell,r)|>\epsilon \Bigr\} \le a_1\exp (-a_2n-a_2\ln n - a_3\ln p ),
\end{equation*}
for some positive constants $a_\ell$ that only depend on~$\epsilon$.  Because of the assumption on the magnitude of~$p$, the \CG{right-hand} side in the above display can be bounded \CG{above} by $a_1p^{-a_4}$.  Thus, taking into account condition~\eqref{c3}, as well as the definition of~$a_j$, we can conclude that, with the exception of the corresponding set of small probability, all~$a_j$ are uniformly bounded.  Moreover, applying Theorem~2.14.9 in \cite{van1996weak} again and taking advantage of~\eqref{c1}, \eqref{c2} and~\eqref{c4}, we can deduce that all~$L_j$ and~$R_j$ are restricted to a uniformly chosen bounded interval, on which all~$\|g_j\|_{\infty}$ and $\|1/g_j\|_{\infty}$ are bounded.

From now on, we focus on all the triples $L\le a\le R$ in the aforementioned bounded interval, for which $\widehat{G}_{\ell, r}^{j}(a)$ is well defined.  Note that for all such triples, and all~$j$, we have $R-a\lesssim P_{j}(a,R)$ and $P_{j}(a,R)\lesssim R-a$. Define $d_{\ell,r}(x)=(r-x)(r-\ell)^{-1}\mathbf{1}({\ell<x<r})$ and note that
\begin{eqnarray*}
|\widehat \mu_{a,R}^{j}-\mu_{a,R}^{j} | &=& \left| \frac{P_{nj}d_{a,R}(R-a)}{P_{nj}(a,R)}- \frac{P_jd_{a,R}(R-a)}{P_j(a,R)} \right|\\
	\\
	&\le&  |P_{nj}d_{a,R}-P_jd_{a,R} |\frac{(R-a)}{P_j(a,R)}+ |P_{nj}(a,R)-P_j(a,R) |\frac{(R-\widehat \mu_{a,R}^{j})}{P_j(a,R)}\\
	\\
	&\lesssim&  |P_{nj}d_{a,R}-P_jd_{a,R} |+ |P_{nj}(a,R)-P_j(a,R) | \equiv  E_1+E_2, \mbox{ say}.
\end{eqnarray*}
Let $h_{\ell, r}(x)=x\mathbf{1}(\ell < x < r)$. Observe that $|P_{nj}h_{a,R}-P_jh_{a,R}|\lesssim E_2$, and define  $D_n=\max_j\sup_{\ell<r}(|P_{nj}h_{\ell, r}-P_jh_{\ell, r}| )+ (|P_{nj}(\ell,r)-P_j(\ell,r)| )$. It follows that
\begin{eqnarray*}
|(\mu_{\ell, r}^j-\mu_{L,a}^j)-(\widehat \mu_{\ell, r}^j-\widehat \mu_{L,a}^j)|
&=&\left|  \frac{P_jh_{L,a}P_j(a,R)+P_jh_{a,R}P_j(L,a)}{P_j(\ell, r)P_j(L,a)}-\frac{P_{nj}h_{L,a}P_{nj}(a,R)+P_{nj}h_{a,R}P_{nj}(L,a)}{P_{nj}(\ell, r)P_{nj}(L,a)} \right| \\ [2mm]
&\lesssim& E_2+ \{(R-a)+E_2 \}D_n.
\end{eqnarray*}
Applying Theorem~2.14.9 in \cite{van1996weak}, together with the union bound, we note that, for each positive~$\epsilon$, we have
\begin{equation*}
\Pr (D_n>\epsilon )\le a_1\exp (-a_2n-a_2\ln n - a_3\ln p ),
\end{equation*}
for some positive constants $a_l$ that only depend on~$\epsilon$.  Because of the assumption on the magnitude of~$p$, the \CG{right-hand} side in the above display can be bounded \CG{above} by $a_1p^{-a_4}$.  Thus, to establish the bound in inequality~\eqref{ap2} we only need to verify that it holds for $E_1$ and $E_2$, uniformly over $L\le a\le R$ in the aforementioned bounded interval.  We will focus on $E_2$, as $E_1$ can be handled with only minor modifications to the argument.  We need to show that there exist positive constants $c_2$ and $c_3$ and a sequence of random variables $M_n$, such that $\mathrm{Pr}\,(M_n>c_3)\lesssim p^{-c2}$ and inequalities
\begin{equation}
\label{ap10}
|P_{nj}(\ell,r)-P_j(\ell,r)| \le \epsilon_1(r-\ell)+\dfrac{\ln (p\vee n)}{n} M_n
\end{equation}
hold for all $j$ and $(\ell,r)$ contained within the bounded interval. Let $M^j_n$ be the infimum of all those values for which equation \eqref{ap10} holds in the \CG{$j$th} coordinate, and define $M_n=\max_{j} M^j_n$.  Recall that we have restricted our attention to a uniformly bounded interval, on which $c_4 =\max_j\|1/g_j\|_{\infty}$ is positive.  Let $\xi_n=\ln(p\vee n)/n$, and write $A_{k,n}^{j}$ for the set of intervals $(\ell,r)$ that lie inside the aforementioned uniformly bounded interval and satisfy inequalities $ (2^{k-1}-1 )\xi_n < P_j(\ell,r) \le 2^k\xi_n$.     In what follows, constants $c_\ell$ are positive and can be chosen independently from $c_3$, $p$, $n$, and $j$.  Observe that
\begin{eqnarray*}
\mathrm{Pr} (M_n>c_3 )& \le& \sum_{j}\mathrm{Pr} (M^j_n>c_3 )\\
	\\
	&\le& \sum_{j}\sum_{k=1}^{\infty} \mathrm{Pr}\,\Bigl\{\exists_{(\ell,r)\in A_{k,n}^{j}} \;  |P_{nj}(\ell, r)-P_j(\ell, r)|>\epsilon_1(r-\ell)+c_3\xi_n\Bigr\} \\
	\\
	&\le&\sum_{j}\sum_{k=1}^{\infty}\mathrm{Pr}\Big[  \sup_{A_{k,n}^j}|P_{nj}(\ell, r)-P_j(\ell, r)|>\xi_n \{c_4\epsilon_1 (2^{k-1}-1 )+c_3 \}\Big].
\end{eqnarray*}
We will bound each summand in the last expression by applying Theorem 2.14.25 in \cite{van1996weak} with  $\mu_n=2^{k/2}\xi_n\sqrt{n}$ and $\sigma^2_{\mathcal{F}}=2^{k}\xi_n$.  Note that~$\mu_n$ is an upper bound on $\sup_{A_{k,n}^j}\sqrt{n} \, |P_{nj}(\ell, r)-P_j(\ell, r)|$, up to some universal multiplicative factors,  as demonstrated in the proof of Lemma~5 in \cite{radchenko2014consistent}.
It follows that
\begin{eqnarray*}
	\mathrm{Pr} (M_n>c_3 )&\le& \sum_{j}\sum_{k=1}^{\infty} \exp (-c_5 c_3 n\xi_n ) \, \exp (-c_62^k ) \\
	&\lesssim&  p \, \exp \{-c_{5}c_3\ln(p\vee n) \} = \exp\{-c_5c_3\ln(p\vee n)+\ln p\}\le p^{-c_6},
\end{eqnarray*}
where $c_{6}$ can be chosen to be positive as long as we take $c_{5}>1/c_3$.

We complete the proof by noting that the remaining bound, \eqref{ap4}, is implied by conditions~\eqref{c2} and~{C2}, the derivations in the proof of Theorem~1 in~\cite{radchenko2014consistent}, and the aforementioned fact that all $L_j$ and $R_j$ lie in a bounded interval with high probability.

The result of Corollary~\ref{cor1} follows from our Theorem~\ref{thm1} and Theorem~1 in \cite{radchenko2014consistent}, due to the fact that the clustering scores, $S_j$, of the signal features are bounded away from zero. \hfill $\Box$

\subsection{Further empirical results}

\subsubsection{Correlated data}
\begin{figure}[!h]
	\centering
	\includegraphics[width=0.5\linewidth]{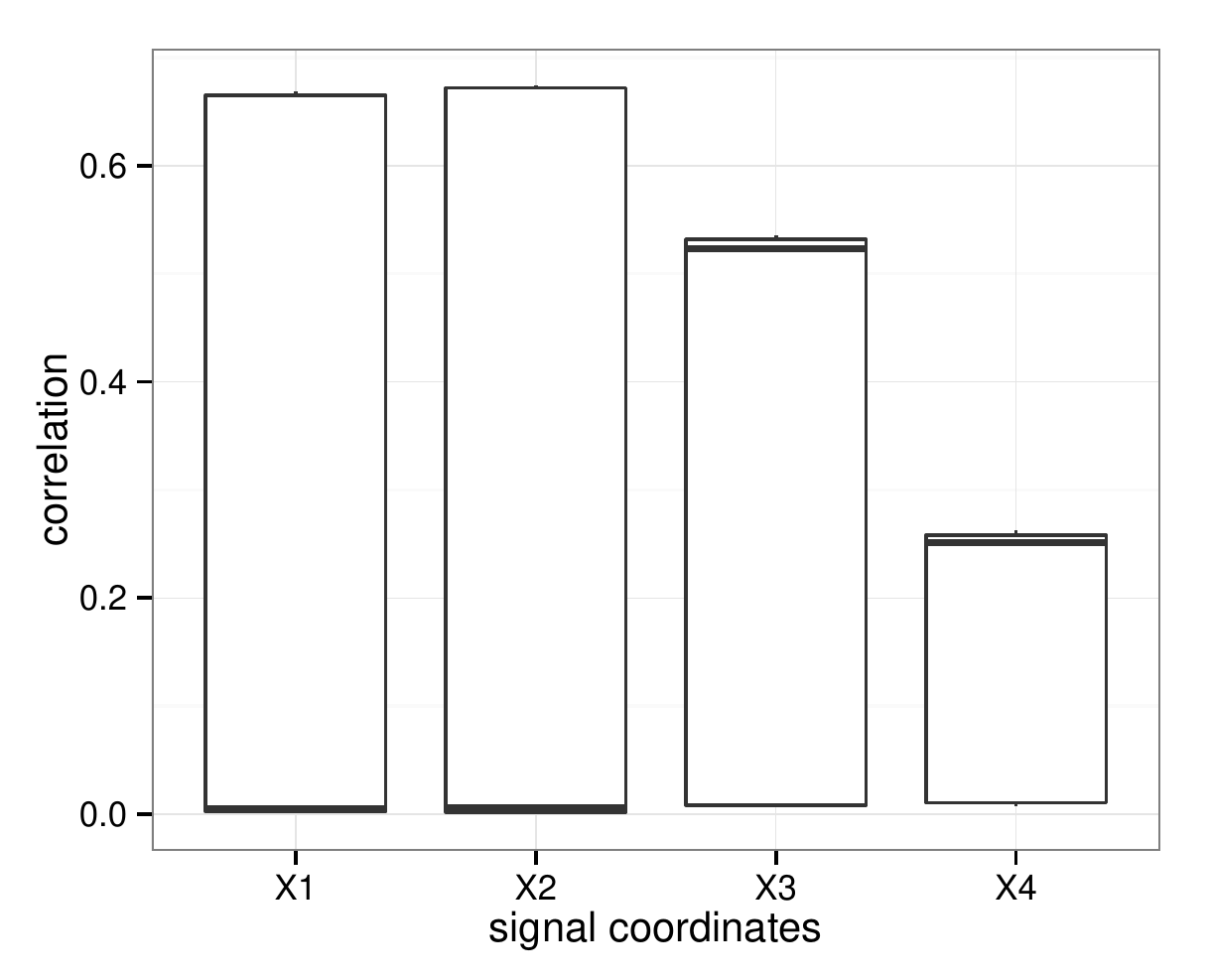}
	\caption{Box plot of the observed average linear correlation between the $p_S$ signal features and the Gaussian noise features. The average is across $10$ repetitions of the simulated data for $n=2000$.}
	\label{corr_ana_2way}
\end{figure}

Here, we assess the screening performance of COSCI when the data has significant correlation between the features. Consider the setup where pairs of features hold cluster information jointly but are un-informative marginally. The design of this experiment is similar to Experiment V presented in Section \ref{sec:combinations}. We set $m=20$, $p_S=4$, $p_N=21$ and $n=2000$. We let the $p_N$ noise features be $\mathcal{N}(0,1)$. For the signal coordinates, we take
\begin{enumerate}
	\item[1.] $(X_1,X_2) \sim  \sum_{i=1}^{2}\mathcal{N}({\mu}_i,\Sigma)/2$
	\item[2.] $X_3 \sim 0.5 \, \mathcal{\beta} (4,6) + 0.5 \,\mathcal{\beta}(7,3)$
	\item[3.] $X_4 \sim 0.5 \, \mathcal{LN}(0.2,0.35) + 0.5 \,\mathcal{N}(4,0.5)$, where
	\begin{eqnarray*}
		{\mu}_1=(0.9,-0.9),\quad {\mu}_2=-{\mu}_1,\quad
		\Sigma =\begin{pmatrix}
			1 & 0.9\\
			0.9 & 1
		\end{pmatrix}.
	\end{eqnarray*}
\end{enumerate}
In this setting features $X_3$ and $X_4$ are bi-modal but $(X_1,X_2)$ are only jointly bi-modal. Note that the effective dimensionality of the data in this example is \CG{$p_S+p_N+m(p_S+p_N)(p_S+p_N-1)/2=6025$}. We introduce dependence between the signal features $(X_1,X_2)$ and the Gaussian noise features $(X_5,\ldots,X_{14})$ through a Gaussian copula with a correlation matrix that has all off-diagonal elements equal to $0.9$. To introduce dependence between the signal features $(X_3,X_4)$ and the Gaussian noise features $(X_{15},\ldots,X_{25})$, we use a multivariate $t$ copula with $2$ degrees of freedom and with a correlation matrix with all off-diagonal elements equal to $0.8$. \CG{(For additional information about copulas, see, e.g., \cite{GF:2007,GN:2012}).} Figure~\ref{corr_ana_2way} presents the distribution of observed average linear correlation between the $p_S$ signal features and the Gaussian noise features. The average is across $10$ repetitions of the simulated data for $n=2000$.
We report the \textit{False Positive} and \textit{False Negative} proportions for the aforementioned simulation study in \CG{Table}~\ref{tab:2way_cn}. COSCI successfully identifies all the signal coordinates. Comparing with \CG{Table}~\ref{tab:2way}, we see that its performance is not affected due to correlation in the data. As expected, higher values of the threshold parameter improve the FP rate, and the benefit of using the data driven approach to select the features is evident. We find that correlation among features does not appear to impact the screening performance of COSCI.

\begin{table}[!h]
\medskip
	\centering
	\caption{False Negatives and False Positive rates for COSCI in studying the impact of correlation. Here, $\mathcal{I}_S=\{1,2,3,4\}$, $p=25$. The numbers in parenthesis are standard errors over $10$ repetitions.}
	\scalebox{0.8}{\begin{tabular}{ccccc}
			\multicolumn{1}{c}{} &       & \multicolumn{2}{c}{$n$ = 2000 / $p$ = 25} & \\
			\midrule
			&       & Avg FN & Avg FP \\
			\midrule
			\multicolumn{1}{c}{} & 0.05  & 0.00 (0.00) & 21.0 (0.00) \\
			\multicolumn{1}{c}{COSCI} & 0.08  & 0.00 (0.00) & 20.7 (0.15) \\
			\multicolumn{1}{c}{with} & 0.1   & 0.00 (0.00) & 19.9 (0.18) \\
			\multicolumn{1}{c}{$\alpha_0$} & 0.12  & 0.00 (0.00) & 18.4 (0.40) \\
			\multicolumn{1}{c}{fixed} & 0.15  & 0.00 (0.00) & 16.0 (0.60) \\
			\multicolumn{1}{c}{} & 0.2   & 0.00 (0.00) & 10.5 (0.65) \\
			\multicolumn{1}{c}{} & 0.25   & 0.00 (0.00) & 4.0 (0.79) \\
			\midrule
			& Data driven & 0.00 (0.00)& 1.00 (0.79)\\
			\bottomrule
		\end{tabular}
		\label{tab:2way_cn}}
\end{table}

\subsubsection{Clustering errors}

In the simulation experiments of Section~\ref{sec:4.1} we evaluated the feature selection performance of COSCI and obtained encouraging results. Here, we check whether COSCI's better feature selection performance also lead to a reduction of the clustering error rates in those experiments.

In \CG{Table}~\ref{simtab6}, we present the clustering errors for Simulation Experiments I and II.
Recall that for Simulation Experiment I, the $p_S=5$ signal features each had two clusters and for Simulation Experiment II, the first five of the $p_S=6$ signal features had two clusters and the sixth one had thee clusters. Thus, the true number of clusters in those experiments is $32$ and $96$ respectively. As the true number of clusters are so large, traditional clustering dissimilarity measures (such as the classification error rates (CER) used in \citet{witten2012framework}) fail to distinguish between good and bad clustering if we compare with respect to all the product clusters. Here, for each signal coordinate we calculate the CER considering the true cluster labeling to be based on that signal only and report the average CER across all the signal coordinates as the clustering error.
Each simulation experiment was repeated  $50$ times and the average clustering error over the $50$ repetitions is reported in the tables.  For Ex.~Mass and COSCI, we use \CG{$k$-means} to perform clustering on the selected features.

\begin{table}[t!]
	\centering
	\caption{Clustering Errors for Simulation Experiment I (left) and II (right).}
	\medskip
	\scalebox{0.7}{\begin{tabular}{ccccc}
			\toprule
			&  & $n = 200$ & $n = 1000$ & $n = 2500$ \\
			\midrule
			\multicolumn{1}{c}{} & 0.05  & 0.178 & 0.242 & 0.178\\
			\multicolumn{1}{c}{COSCI} & 0.08  & 0.238 & 0.229 & 0.165\\
			\multicolumn{1}{c}{with} & 0.1   & 0.264 & 0.229 & 0.138 \\
			\multicolumn{1}{c}{$\alpha_0$} & 0.12  &  0.305 & 0.222 & 0.122\\
			\multicolumn{1}{c}{fixed} & 0.15  & 0.338 & 0.211 & 0.114\\
			\multicolumn{1}{c}{} & 0.2   & 0.344 & 0.183 & 0.121\\
			\midrule
			\multicolumn{1}{c}{} & Data driven &  0.195 & 0.172 & 0.133\\
			\midrule
			\multicolumn{1}{c}{} &   SpKM & 0.484 & 0.46 & 0.47 \\
			\multicolumn{1}{c}{Other} &  SpHC & 0.495 & 0.491 & 0.494\\
			\multicolumn{1}{c}{methods} &   SAS & 0.495 & 0.494 & 0.478 \\
			\multicolumn{1}{c}{} &  Ex. Mass & 0.234 & 0.203 & 0.203\\
			\multicolumn{1}{c}{} &  IF-PCA & 0.478 & 0.484 & 0.488\\
			\bottomrule
	\end{tabular}}%
	\;\;\;\;\;\;
	\scalebox{0.7}{\begin{tabular}{ccccc}
			\toprule
			&  & $n = 200$ & $n = 1000$ & $n = 2500$ \\
			\midrule
			\multicolumn{1}{c}{} & 0.05  & 0.279 & 0.303 & 0.221\\
			\multicolumn{1}{c}{COSCI} & 0.08  &0.3 & 0.28 & 0.211\\
			\multicolumn{1}{c}{with} & 0.1   & 0.316 & 0.265 & 0.185 \\
			\multicolumn{1}{c}{$\alpha_0$} & 0.12  & 0.341 & 0.256 & 0.176\\
			\multicolumn{1}{c}{fixed} & 0.15  & 0.370 & 0.245 & 0.175\\
			\multicolumn{1}{c}{} & 0.2   & 0.398 & 0.231 & 0.184\\
			\midrule
			\multicolumn{1}{c}{} & Data driven &  0.248 & 0.220 & 0.179\\
			\midrule
			\multicolumn{1}{c}{} &   SpKM & 0.495 & 0.486 & 0.495 \\
			\multicolumn{1}{c}{Other} &  SpHC & 0.509 & 0.509 & --- \\
			\multicolumn{1}{c}{methods} &   SAS & 0.493 & 0.491 & 0.495 \\
			\multicolumn{1}{c}{} &  Ex. Mass & 0.283 & 0.253 & 0.252\\
			\multicolumn{1}{c}{} &  IF-PCA & 0.475 & 0.485 & 0.486\\
			\bottomrule
		\end{tabular}
		\label{simtab6}}
\end{table}

In \CG{Table}~\ref{simtab6}, it was seen that COSCI coupled with the data driven approach continues to provide the best CER amongst all the competing methods. IF-PCA, SAS, SpKM and SpHC exhibit very high CER in these settings even though their screening performance, especially that of SAS and IF-PCA, was competitive (see Tables \ref{simtab1}--\ref{simtab2} in Section~\ref{sec:4.1}). COSCI and Ex.~Mass return far better CER and as is expected, with $n$ small and increasing $\alpha_0$, COSCI screening begins to miss the signal features and therefore returns a higher CER. Similar characteristics in the clustering efficacies were also observed across the other simulation scenarios of Section~\ref{sec:4.1}.

\subsubsection{COSCI Analysis on Cardio and RNA Seq Datasets}
\label{sec:appc2}
\begin{figure}[H]
	\centering
	\includegraphics[width=34pc,height=20pc]{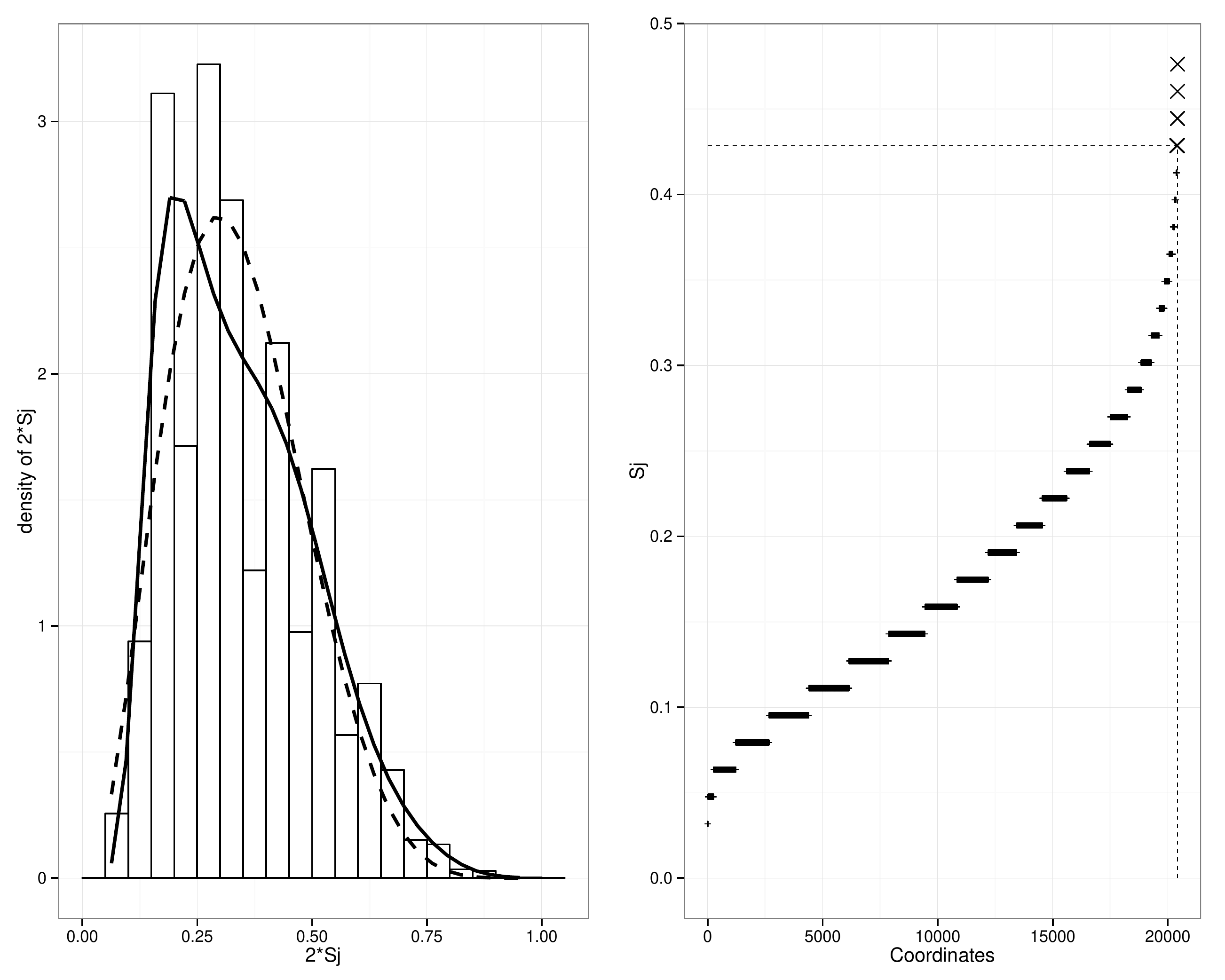}
	\caption{\textsc{Cardio Data.} Left: $\widehat{\pi}_{0}\widehat{f_{0}}(\psi_j)$ in black dashed line and $\widehat{f}(\psi_j)$ in black solid line for $j \in \{1,\ldots,\mbox{20426}\}$. Right: Distribution of $S_j$. The 33 selected features are marked as $\times$. The dashed horizontal line is $\widehat{\alpha}_0=0.428$.}
	\label{cardiof}
\end{figure}%
\begin{figure}[!h]
	\centering
	\includegraphics[width=34pc,height=20pc]{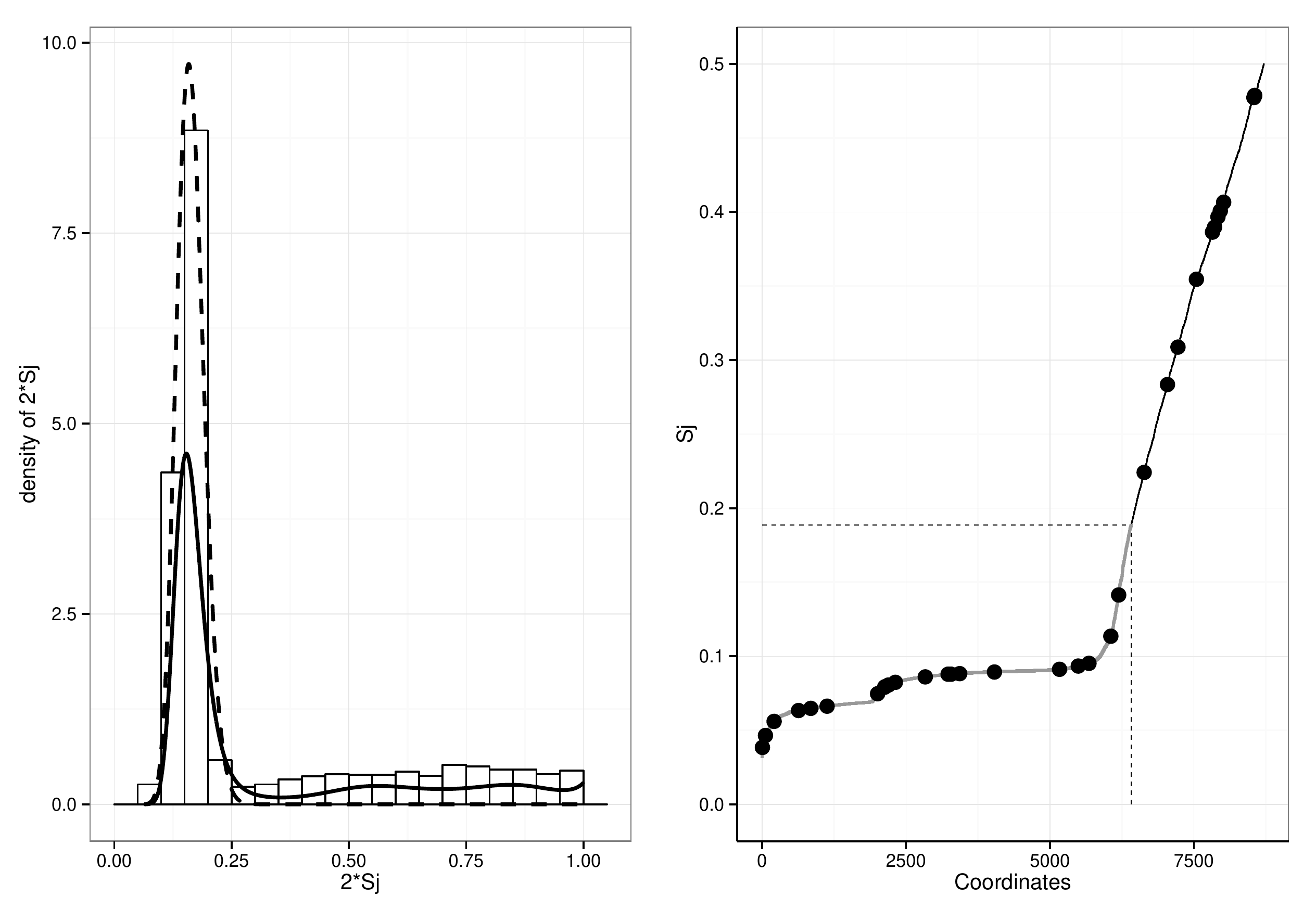}
	\caption{\textsc{RNASeq Data.} Left: $\widehat{\pi}_{0}\widehat{f_{0}}(\psi_j)$ inblack dashed line and $\widehat{f}(\psi_j)$ in black solid line for $j \in \{1,\ldots,8716\}$. Right: Distribution of $S_j$. The dashed horizontal line is $\widehat{\alpha}_0=0.188$. The 2304 selected features are in black and are above the dashed horizontal line. The black dots are the 33 lineage markers.}
	\label{rnaf}
\end{figure}%
\subsubsection{More real data examples}\label{sec:appc1}
In Table \ref{realcer2}, we present more real data examples to assess the performance of COSCI. We follow the theme presented in \CG{Section}~\ref{sec:realdata} and overlay COSCI with $k$-means, Sparse $k$-means and IF-PCA for each of the eleven datasets considered below. The first ten data sets are sourced from Jiashun Jin's webpage (see: \url{http://www.stat.cmu.edu/~jiashun/Research/software/GenomicsData/} and Table 1 in \cite{jin2016influential} for more information on these data sets.). The last data set is available on Brad Efron's webpage (see: \url{http://statweb.stanford.edu/~ckirby/brad/LSI/datasets-and-programs/datasets.html}).

\begin{table}[b!]
	\centering
	\caption{Minimum CER of the competing methods \CG{along with} the method name. For COSCI, the smallest CER observed for COSCI + ``clustering method'' is shown where ``clustering method'' includes the \CG{five} competing methods in Table \ref{realcer1}. Number of selected features are reported after the `/' symbol.}
	\scalebox{0.8}{\begin{tabular}{llccccccc}
			\toprule
			&       &       &       &       & \multicolumn{2}{c}{Min. CER} & \multicolumn{2}{c}{COSCI}\\
			\midrule
			\multicolumn{1}{l}{Data Set} & Source & $n$     & $p$     & $K$     & Method & CER   & Min. CER   & Method \\
			\midrule
			ProstateCancer & Singh et al. (2002)      & 102   & 6033  & 2     & IF-PCA & 0.477 / 1551 & 0.498 / 178 & $k$-means \\
			Lymphoma & Alizadeh et al. (2000)      & 62    & 4026  & 3     & IF-PCA & 0.103 / 42 & 0.285 / 22 & IF-PCA \\
			Brain &  Pomeroy et al. (2002)     & 42    & 5597  & 5     & IF-PCA & 0.159 / 453 & 0.131 / 323 & IF-PCA \\
			Colon &  Alon et al. (1999)     & 62    & 2000  & 2     & IF-PCA & 0.490 / 25 & 0.444 / 3 & $k$-means \\
			SRBCT &  Kahn (2001)     & 63    & 2308  & 4     & SAS   & 0.327 / 246 & 0.350 / 84 & $k$-means \\
			Leukemia &  Golub et al. (1999)     & 72    & 3571  & 2     & IF-PCA & 0.131 / 213 & 0.430 / 112 & IF-PCA \\
			SuCancer &  Su et al. (2001)     & 174   & 7909  & 2     & SAS   & 0.501 / 526 & 0.500 / 211 & $k$-means \\
			LungCancer(1) & Gordon et al. (2002)      & 181   & 12533 & 2     & IF-PCA & 0.064 / 251 & 0.206 / 420 & $k$-means \\
			LungCancer(2) &  Bhattacharjee et al. (2001)     & 203   & 12600 & 2     & IF-PCA & 0.341 / 418 & 0.502 / 260 & $k$-means \\
			BreastCancer  & Wang et al. (2005)      & 276   & 22215 & 2     & IF-PCA & 0.484 / 721 & 0.492 / 1195 & $k$-means \\
			Michigan &   Subramanium et al. (2005)    & 86    & 5217  & 2     & Ex. Mass & 0.427 / 4 & 0.479 /51 & $k$-means\\
			\bottomrule
		\end{tabular}
		\label{realcer2}}
\end{table}

IF-PCA returns a significantly smaller error rate on the Lymphoma, Leukemia and Lung Cancer datasets but barring these four datasets (Lymphoma, Leukemia, Lung Cancer (1) and Lung Cancer (2)), feature screening by COSCI returns comparable and often better error rates using fewer features when compared to the best competing method. In the datasets of Breast Cancer and Michigan COSCI selects more features than the best competing method but returns a comparable error rate. Detailed introspection on the four datasets where IF-PCA performs better than COSCI reveals that the overlap between these two methods in the screening step is minimal here. COSCI fails to pick up those signal features that pass through its merge size filter and are statistically far away from a Gaussian distribution. \CG{In contrast,} IF-PCA precisely picks up these features which happen to be the best features to perform clustering in these four datasets. In simulation Experiments III--IV and in the real datasets considered in Section \ref{sec:realdata}, the setting is, however, exactly opposite where the best signal features are not uni-modal and the noise features are not always Normally distributed. In these settings, COSCI prospers and by far returns the best screening performance than all the other competing methods considered.

\subsection{Algorithm details}
\subsubsection{Detailed description of Algorithm~1}
\label{sec:appe}
\scalebox{0.85}{
	\begin{algorithm}[H]
		\caption{COSCI procedure for feature screening.} 
		\textit{input}: data matrix ${X}^{n\times p}$ and tuning parameter $\alpha_0$
		
		\textit{output}: merge sizes $\{S_j\}_{1}^{p}$ for $p$ features and feature set $\mathcal{\widehat{I}}_{S}$
		\begin{tabbing}
			\enspace FOR each $j\in\{1,2,\ldots,p\}$ \\
			\qquad INITIALIZE;\\
			\hspace{20pt}$k$ = number of clusters = $n$\\
			\hspace{20pt}sort $x =\{x_1,\ldots,x_n\}$ in ascending order\\
			\hspace{20pt}assign cluster mean: $a_i = x_i$ for $i \in\{ 1,\ldots,n\}$\\
			\hspace{20pt}assign cluster size: $s_i = 1$, $i \in \{1,\ldots,n\}$\\
			\hspace{20pt}assign cluster membership indices of $x$: $I(x) = \{1,\ldots,n\}$\\
			\qquad WHILE $k>1$\\[5pt]
			\qquad\textit{Convex Merging Algorithm}\\
			\qquad\tcc{1. Find the consecutive adjacent centroid distances}\\
			\qquad	$d(r,r + 1)\leftarrow (a_{r+1}-a_{r})/(s_r + s_{r+1})$\\
			\qquad\tcc{2. Find clusters with minimum merging distance}\\
			\qquad	$r^{\star}\leftarrow\arg\min_{1\le r\le k-1}d(r,r + 1)$\\[5pt]
			\qquad\textit{Merge Sizes}\\
			\qquad\tcc{3. Determine merge size $\alpha_{i}^{j}$}\\
			\qquad	$\alpha_{i}^{j}=n^{-1}\min (s_{r^{\star}},s_{r^{\star}+1})$, $i=$\textit{merge index}\\[5pt]
			\qquad\textit{Screening the Merges}\\
			\qquad\tcc{4. Obtain mass after merge $m_i$}\\
			\qquad	$m_i =n^{-1}\left(s_{r^{\star}}+s_{r^{\star}+1}\right)$\\
			\qquad\qquad IF $m_i<0.5$ then $\alpha_{i}^{j}=0$\\[5pt]
			\qquad\textit{Prepare for the next iteration}\\
			\qquad\tcc{5. Merge $r^{\star},r^{\star}+1$ clusters}\\
			\qquad	$a_{r^{\star}}\leftarrow \left(s_{r^{\star}}a_{r^{\star}}+s_{r^{\star}+1}a_{r^{\star}+1}\right)/(s_{r^{\star}}+s_{r^{\star}+1})$\\
			\qquad	$s_{r^{\star}}\leftarrow s_{r^{\star}}+s_{r^{\star}+1}$\\
			\qquad\tcc{6. Reduce number of clusters}\\
			\qquad	$k\leftarrow k-1$\\
			\qquad\tcc{7. Change cluster and member indices}\\
			\qquad FOR $\ell$ in $(r^{\star}+1):k$, $s_\ell \leftarrow s_{\ell+1},a_l\leftarrow a_{\ell+1}$\\
			\qquad FOR ALL $I(x)>r^{\star}$: $I(x)=I(x)-1$\\[5pt]
			\quad\textit{Store max merge sizes}\\
			\quad store $S_j=\max_{1\le i\le (n-1)}\alpha_{i}^{j}$
		\end{tabbing}%
		\textit{Feature screening}\\
		Choose $\mathcal{\widehat{I}}_{S} = \{j:S_j\ge\alpha_{0}\}$\\
\end{algorithm}}

\subsubsection{Estimation of Hyperparameters ---  simulation based}\label{sec:appd}

\begin{table}[H]
	\scriptsize
	\centering
	\caption{This table reports the $\%$ of cases where COSCI detects clusters across varying sample sizes $n$ and thresholds $\alpha_0$.}
	\scalebox{1}{\begin{tabular}{crrrrrrrrr}
			\toprule
			{} & {} &  &  & $\alpha_0$ &  &  &\\
			\midrule
			Unimodal density & $n$ & 0.01 & 0.02 & 0.05 & 0.1 & 0.15 & 0.2 & 0.25 \\
			\midrule
			{} & 100 & 100 & 100 & 99 & 74 & 47 & 28 & 12 \\
			& 500 & 100 & 100 & 67 & 33 & 17 & 11 & 7 \\
			& 1000 & 100 & 98 & 49 & 22 & 13 & 6 & 2 \\
			$\mathcal{N}(0,1)$ & 2000 & 100 & 82 & 21 & 10 & 3 & 0 & 0 \\
			& 5000 & 94 & 38 & 3 & 1 & 1 & 0 & 0 \\
			& 10000 & 47 & 6 & 0 & 0 & 0 & 0 & 0 \\
			\midrule
			{}& 100 & 100 & 100 & 59 & 19 & 11 & 3 & 1 \\
			\multicolumn{1}{l}{} & 500 & 99 & 50 & 4 & 0 & 0 & 0 & 0 \\
			\multicolumn{1}{l}{} & 1000 & 74 & 14 & 0 & 0 & 0 & 0 & 0 \\
			Student $t_{(1)}$  & 2000 & 27 & 0 & 0 & 0 & 0 & 0 & 0 \\
			\multicolumn{1}{l}{} & 5000 & 0 & 0 & 0 & 0 & 0 & 0 & 0 \\
			\multicolumn{1}{l}{} & 10000 & 0 & 0 & 0 & 0 & 0 & 0 & 0 \\
			\midrule
			{}& 100 & 100 & 100 & 96 & 59 & 33 & 14 & 7\\
			\multicolumn{1}{l}{} & 500 & 100 & 99 & 47 & 10 & 2 & 2 & 0 \\
			\multicolumn{1}{l}{} & 1000 & 100 & 84 & 7 & 2 & 0 & 0 & 0 \\
			$\mathcal{E}(1)$  & 2000 & 95 & 33 & 0 & 0 & 0 & 0 & 0 \\
			\multicolumn{1}{l}{} & 5000 & 35 & 0 & 0 & 0 & 0 & 0 & 0 \\
			\multicolumn{1}{l}{} & 10000 & 2 & 0 & 0 & 0 & 0 & 0 & 0 \\
			\midrule
			{}& 100 & 100 & 99 & 52 & 18 & 10 & 2 & 1 \\
			\multicolumn{1}{l}{} & 500 & 99 & 62 & 7 & 1 & 0 & 0 & 0 \\
			\multicolumn{1}{l}{} & 1000 & 82 & 17 & 0 & 0 & 0 & 0 & 0 \\
			Cauchy  & 2000 & 36 & 3 & 0 & 0 & 0 & 0 & 0 \\
			\multicolumn{1}{l}{} & 5000 & 0 & 0 & 0 & 0 & 0 & 0 & 0 \\
			\multicolumn{1}{l}{} & 10000 & 0 & 0 & 0 & 0 & 0 & 0 & 0 \\
			\midrule
			{}& 100 & 100 & 100 & 75 & 29 & 13 & 2 & 1 \\
			\multicolumn{1}{l}{} & 500 & 100 & 81 & 11 & 1 & 0 & 0 & 0 \\
			\multicolumn{1}{l}{} & 1000 & 98 & 40 & 1 & 0 & 0 & 0 & 0 \\
			$\mathcal{L}(1)$  & 2000 & 71 & 8 & 0 & 0 & 0 & 0 & 0 \\
			\multicolumn{1}{l}{} & 5000 & 4 & 0 & 0 & 0 & 0 & 0 & 0 \\
			\multicolumn{1}{l}{} & 10000 & 0 & 0 & 0 & 0 & 0 & 0 & 0 \\
			\midrule
			& 100 & 100 & 100 & 80 & 28 & 16 & 7 & 2 \\
			GEV with & 500 & 100 & 82 & 15 & 3 & 0 & 0 & 0 \\
			shape & 1000 & 99 & 48 & 3 & 0 & 0 & 0 & 0 \\
			parameter  & 2000 & 68 & 12 & 0 & 0 & 0 & 0 & 0 \\
			= 0.8 & 5000 & 13 & 0 & 0 & 0 & 0 & 0 & 0 \\
			\multicolumn{1}{l}{} & 10000 & 0 & 0 & 0 & 0 & 0 & 0 & 0 \\
			\midrule
			& 100 & 100 & 100 & 99 & 78 & 50 & 32 & 19 \\
			$\mathcal{B} (1,3)$& 500 & 100 & 100 & 74 & 28 & 16 & 11 & 2 \\
			& 1000 & 100 & 98 & 37 & 5 & 3 & 1 & 0 \\
			& 2000 & 99 & 80 & 14 & 2 & 0 & 0 & 0 \\
			& 5000 & 88 & 24 & 2 & 0 & 0 & 0 & 0 \\
			\multicolumn{1}{l}{} & 10000 & 43 & 1 & 0 & 0 & 0 & 0 & 0 \\
			\midrule
			& 100 & 100 & 100 & 99 & 84 & 54 & 33 & 22 \\
			Triangle distbn.  & 500 & 100 & 100 & 79 & 34 & 15 & 7 & 0 \\
			$\in [0,1]$ & 1000 & 100 & 98 & 42 & 11 & 4 & 4 & 1 \\
			with   & 2000 & 100 & 88 & 28 & 5 & 1 & 0 & 0 \\
			mode at 0.8 & 5000 & 98 & 40 & 3 & 0 & 0 & 0 & 0 \\
			\multicolumn{1}{l}{} & 10000 & 68 & 11 & 0 & 0 & 0 & 0 & 0\\
			\bottomrule
		\end{tabular}
		\label{tabsim}}
\end{table}

\subsubsection{Two-stage approach to signal screening}\label{appb}

We briefly describe the \CG{two-stage} approach to signal screening introduced in \citep{tony2016optimal}. We start with Eq.~ \eqref{fdr} where the estimated fdr has been obtained as
\begin{equation*}
T_j = \widehat{\pi_0}\widehat{f_{0}}(\psi_j)/\widehat{f}(\psi_j)
\end{equation*}
Order the estimated fdr's $T_j$ from smallest to largest so that $T_{(1)}\le\cdots\le T_{(p)}$. In the first stage, we estimate \CG{Stage}~1 screening cutoff $k_s$ by
\begin{equation*}
k_s=\min\Bigl\{j:\sum_{i=p}^{j}\left(1-T_{(i)}\right)\le p(1-\widehat{\pi}_0)\delta_p\Bigr\}
\end{equation*}
with $\delta_p=\min\{ p,(\ln p)^{-1}\}$ and select the features as
\begin{equation*}
\mathcal{S}_s= \{j:T_j\le T_{(k_s)} \}= \{T_{(1)},\ldots,T_{(k_s)} \}
\end{equation*}
In the second stage, the so-called \textit{Discovery stage}, we take the \CG{Stage}~1 screened fdr's $ l\{T_j:j\in\mathcal{S}_s \}$ and estimate the \CG{Stage}~2 screening cutoff $k_d$ by
\begin{equation*}
k_d=\max\left\{1\le j\le k_s:\dfrac{1}{j}\sum_{i=1}^{j}T_{(j)}\le\delta_p\right\}
\end{equation*}
and select the features as
\begin{equation*}
\mathcal{\widehat{I}}_S= \{j\in\mathcal{S}_s:T_j\le T_{(k_d)} \}
\end{equation*}
Proposition 2 in \cite{tony2016optimal} guarantees that in \CG{Stage}~1, $\mathcal{S}_s$ is the largest subset such that the missed discovery rate (MDR) is controlled at level $\delta_p$ while in \CG{Stage}~2, $\mathcal{\widehat{I}}_{S}$ is the smallest subset such that the false positive rate (FPR) is controlled at level $\delta_p$.

\subsubsection{Analysis of the population procedure for Gaussian mixtures}\label{appendix:normals}

In Table~\ref{table1}, which is an adapted version of Table~1 in the supplementary material for~\cite{radchenko2014consistent}, we document the behavior of the population clustering procedure for a wide variety of mixtures of two Gaussian distributions on the real line. For \CG{seven} different levels of separation between the two means we consider \CG{nine} different mixing proportions, from the symmetric case of 50:50 mixing to the highly skewed 10:90 mixing.  The behavior of the population splitting procedure in other cases can be interpolated using continuity arguments. In all of the scenarios, the population procedure identifies either two clusters or one.  The latter happens only in the settings where the separation between the sub-populations, or the size of one sub-population, is very small.  We also present the location of the split point, $s^*$ (``NO'' denotes the cases where no splits are detected, and, thus, only one cluster is identified), the local minimum of the density, $\dmin$, and the split point minimizing the expected misclassification error, $s_{\text{MC}}$.  Finally, we report, under \textit{Excess MCE},  how much the misclassification error of the population clustering procedure exceeds that of the $s_{\text{MC}}$ based oracle rule.
\begin{table}
\centering
\caption{\small{Finding the population splits for $2$--normal mixtures: $p_1 \, \mathcal{N}(\mu_1,1) +p_2\, \mathcal{N}(\mu_2,1)$.}}\label{table1}
\scalebox{0.73}{
\begin{tabular}{crrrrcccc}
 CASE & $p_1$ & $p_2$ & $\mu_1$ & $\mu_2$ &  $\dmin$ & $\sstar$ & $s_{\text{MC}}$ & \;\; Excess MCE \\ 
  \toprule
\multirow{9}{*}{$\bm{|\mu_2-\mu_1| = 9}$} 
  & 0.50 & 0.50 & --4.50 & 4.50 &  0.00 & 0.00  &  0.00 & $0.000$\\ 
  & 0.45 & 0.55 & --4.50 & 4.50 &  --0.02 & --0.45 &  --0.02 & $0.000$\\ 
  & 0.40 & 0.60 & --4.50 & 4.50 &  --0.05 & --0.90 &  --0.04 & $0.000$\\ 
  & 0.35 & 0.65 & --4.50 & 4.50 &  --0.07 & --1.36 &  --0.07 & $0.000$ \\ 
  & 0.30 & 0.70 & --4.50 & 4.50 &  --0.10 & --1.82 &  --0.09 & $0.001$ \\ 
  & 0.25 & 0.75 & --4.50 & 4.50 &  --0.13 & --2.31 & --0.12 & $0.004$\\ 
  & 0.20 & 0.80 & --4.50 & 4.50 &  --0.16 & --2.90 &  --0.15 & $0.011$\\ 
  & 0.15 & 0.85 & --4.50 & 4.50 & --0.20 & --3.82 &  --0.19 & $0.037$\\ 
  & 0.10 & 0.90 & --4.50 & 4.50 &  --0.26 & NO &    --0.24 & $0.100$\\ 
   \midrule
   \multirow{9}{*}{$\bm{|\mu_2-\mu_1| = 8}$}
   & 0.50 & 0.50 & --4.00 & 4.00 & 0.00 & 0.00  &  0.00 & $0.000$ \\ 
   & 0.45 & 0.55 & --4.00 & 4.00 & --0.03 & --0.40 & --0.03 & $0.000$\\ 
   & 0.40 & 0.60 & --4.00 & 4.00  & --0.05 & --0.80  & --0.05 & $0.000$\\ 
   & 0.35 & 0.65 & --4.00 & 4.00  & --0.08 & --1.22  &  --0.08 & $0.001$\\ 
   & 0.30 & 0.70 & --4.00 & 4.00  & --0.11 & --1.64  &  --0.11 & $0.003$\\ 
   & 0.25 & 0.75 & --4.00 & 4.00  & --0.15 & --2.12  &  --0.14 & $0.008$\\ 
   & 0.20 & 0.80 & --4.00 & 4.00 &  --0.18 & --2.72  &  --0.17 & $0.020$\\ 
   & 0.15 & 0.85 & --4.00 & 4.00 &  --0.23 & NO &  --0.22 & $0.150$\\ 
   & 0.10 & 0.90 & --4.00 & 4.00 & --0.29 & NO &    --0.28 & $0.100$ \\ 
   \midrule
   \multirow{9}{*}{$\bm{|\mu_2-\mu_1| = 7}$}
   & 0.50 & 0.50 & --3.50 & 3.50 & 0.00 & 0.00 &  0.00 & $0.000$ \\ 
   & 0.45 & 0.55 & --3.50 & 3.50  & --0.03 & --0.35  & --0.03 & $0.000$\\ 
   & 0.40 & 0.60 & --3.50 & 3.50 & --0.06 & --0.71  & --0.06 & $0.001$\\ 
   & 0.35 & 0.65 & --3.50 & 3.50  & --0.10 & --1.09  & --0.09 & $0.003$\\ 
   & 0.30 & 0.70 & --3.50 & 3.50  & --0.13 & --1.49   & --0.12 & $0.007$ \\ 
   & 0.25 & 0.75 & --3.50 & 3.50  & --0.17 & --1.97  & --0.16 & $0.016$\\ 
   & 0.20 & 0.80 & --3.50 & 3.50 & --0.22 & --2.66   & --0.20 & $0.040$\\ 
   & 0.15 & 0.85 & --3.50 & 3.50  & --0.27 & NO & --0.25 & $0.150$\\ 
   & 0.10 & 0.90 & --3.50 & 3.50  & --0.34 & NO & --0.31 & $0.100$\\ 
   \midrule
      \multirow{9}{*}{$\bm{|\mu_2-\mu_1| = 6}$}
      & 0.50 & 0.50 & --3.00 & 3.00  & 0.00 & 0.00  &  0.00 & 0.000 \\ 
      & 0.45 & 0.55 & --3.00 & 3.00  & --0.04 & --0.32  & --0.03 &0.001\\ 
      & 0.40 & 0.60 & --3.00 & 3.00  & --0.08 & --0.64  &--0.07 & 0.004\\ 
      & 0.35 & 0.65 & --3.00 & 3.00  & --0.12 & --0.99  &--0.10 & 0.008\\ 
      & 0.30 & 0.70 & --3.00 & 3.00 & --0.16 & --1.39  & --0.14 &0.016\\ 
      & 0.25 & 0.75 & --3.00 & 3.00  & --0.21 & --1.91  &--0.18 & 0.034\\ 
      & 0.20 & 0.80 & --3.00 & 3.00  & --0.26 & NO &  --0.23 & 0.200\\ 
      & 0.15 & 0.85 & --3.00 & 3.00  & --0.33 & NO &  --0.29 & 0.150\\ 
      & 0.10 & 0.90 & --3.00 & 3.00 & --0.41 & NO & --0.37 & 0.100\\ 
      \hline
      \hline
      \multirow{9}{*}{$\bm{|\mu_2-\mu_1| = 5}$} 
      & 0.50 & 0.50 & --2.50 & 2.50 &  0.00 & 0.00   & 0.00 & 0.000 \\ 
      & 0.45 & 0.55 & --2.50 & 2.50 & --0.05 & --0.30   &--0.04 & 0.005\\ 
      & 0.40 & 0.60 & --2.50 & 2.50 &  --0.10 & --0.61   &--0.08 & 0.011\\ 
      & 0.35 & 0.65 & --2.50 & 2.50 & --0.15 & --0.96   &--0.12 & 0.021\\ 
      & 0.30 & 0.70 & --2.50 & 2.50 &  --0.20 & --1.41  &--0.17 & 0.041\\ 
      & 0.25 & 0.75 & --2.50 & 2.50 & --0.26 & NO &  --0.22 & 0.250\\ 
      & 0.20 & 0.80 & --2.50 & 2.50 &  --0.33 & NO   &--0.28 & 0.200\\ 
      & 0.15 & 0.85 & --2.50 & 2.50 & --0.41 & NO  & --0.35 & 0.149\\ 
      & 0.10 & 0.90 & --2.50 & 2.50 & --0.53 & NO  & --0.44 & 0.100\\ 
      \midrule
      \multirow{9}{*}{$\bm{|\mu_2-\mu_1| = 4}$} 
      & 0.50 & 0.50 & --2.00 & 2.00 & 0.00 & 0.00  &  0.00 & 0.000 \\ 
      & 0.45 & 0.55 & --2.00 & 2.00 & --0.07 & --0.32  &  --0.05 &  0.015\\ 
      & 0.40 & 0.60 & --2.00 & 2.00 & --0.14 & --0.67  & --0.10 &0.034\\ 
      & 0.35 & 0.65 & --2.00 & 2.00 &--0.21 & --1.12  & --0.15 & 0.065\\ 
      & 0.30 & 0.70 & --2.00 & 2.00 & --0.28 & NO &  --0.21 & 0.298\\ 
      & 0.25 & 0.75 & --2.00 & 2.00 & --0.37 & NO &  --0.28 & 0.248\\ 
      & 0.20 & 0.80 & --2.00 & 2.00 & --0.47 & NO &  --0.35 & 0.198\\ 
      & 0.15 & 0.85 & --2.00 & 2.00 & --0.58 & NO &  --0.43 & 0.148\\ 
      & 0.10 & 0.90 & --2.00 & 2.00 & --0.74 & NO &   --0.55 & 0.097\\ 
      \midrule
      \multirow{9}{*}{$\bm{|\mu_2-\mu_1| = 3}$}
      & 0.50 & 0.50 & --1.50 & 1.50  & 0.00 & 0.00  & 0.00 & 0.000\\ 
      & 0.45 & 0.55 & --1.50 & 1.50  & --0.12 & --0.50  & --0.07 & 0.057 \\ 
      & 0.40 & 0.60 & --1.50 & 1.50  & --0.24& NO   & --0.14 & 0.396\\ 
      & 0.35 & 0.65 & --1.50 & 1.50 & --0.38 & NO   & --0.21 & 0.344\\ 
      & 0.30 & 0.70 & --1.50 & 1.50 & --0.53 & NO  & --0.28 & 0.293\\ 
      & 0.25 & 0.75 & --1.50 & 1.50 & --0.71 & NO  & --0.37 & 0.241\\ 
      & 0.20 & 0.80 & --1.50 & 1.50 &--1.50 & NO  & --0.46 & 0.190\\ 
      & 0.15 & 0.85 & --1.50 & 1.50 & --1.50 & NO  & --0.58 & 0.139\\ 
      & 0.10 & 0.90 & --1.50 & 1.50 & --1.50 & NO  & --0.73 & 0.090\\ 
      \bottomrule
\end{tabular}}
\end{table}

\bigskip
\noindent
\textbf{Acknowledgments}.
We would like to thank the Editor, the Associate Editor and two anonymous referees for many helpful suggestions that improved the paper. We thank Sara Garcia and Gregory Giecold for sharing the RNASeq data and ECLAIR source codes with us. Mukherjee's research was partially supported by the Zumberge individual award from the University of Southern California's James H. Zumberge Faculty Research and Innovation Fund.

\section*{Supplementary material}
\label{SM}
The R code and the data sets used in this paper can be downloaded from the following link - \newline \url{http://github.com/trambakbanerjee/COSCI}.


\end{document}